\documentclass[aps,floatfix]{revtex4}

\usepackage{hyperref,hypcap,ae,tikz,color}

\usepackage{graphicx}
\usepackage{bm}
\usepackage{amsmath}
\usepackage[T1]{fontenc}
\usepackage{xcolor}
\usepackage{lmodern}
\usepackage[caption=false]{subfig}

\hypersetup{
	colorlinks=true,
	citecolor=blue,
    linkcolor=magenta,
	}

\begin{document}

\title{From entropy to gravitational entropy}

\author{\bf{Sarbari Guha}}

\affiliation{\bf Department of Physics, St.Xavier's College (Autonomous), Kolkata 700016, India}

\maketitle
\section*{Abstract}
The concept of entropy forms the backbone of the principles of thermodynamics. It was R.C. Tolman who initiated a correlation between gravity and thermodynamics. The development of black hole thermodynamics and the consideration of the generalized second law of thermodynamics led to Penrose's conjecture that the Weyl tensor should serve as a measure of the entropy of the free gravitational field. This entropy, better known as gravitational entropy, reflects the degrees of freedom associated with the free gravitational field. The proposition of gravitational entropy provides a justification for the low entropy of the initial Universe. It was necessary to associate an entropy function with the dynamics of the free gravitational field even at the time of the big bang, in order that a gravity-dominated evolution of the Universe may preserve the second law of thermodynamics. It also places the concept of black hole entropy in a proper context, which appears as a particular case of the entropy of the free gravitational field. However, a self-consistent notion of gravitational entropy in the context of cosmological structure formation has eluded us till today. Various proposals have been put forward, initially based on Penrose's Weyl Curvature Hypothesis, and subsequently modified to fit the needs of specific geometries and matter distributions. Such proposals were basically geometric in nature. Just a decade back a new definition of gravitational entropy was proposed from the considerations of the relativistic Gibbs equation and based on the square root of the Bel-Robinson tensor, the simplest divergence-free tensor derived from the Weyl tensor. However, even this proposal is valid only for a restricted class of spacetimes. A complete description of gravitational entropy encompassing black hole physics and cosmological dynamics is yet to emerge. In this article, we will try to gather an overview of the concept of gravitational entropy beginning with the initial attempts to correlate gravity with thermodynamics and follow it up with the development of the various proposals of gravitational entropy within the framework of general relativity.

\bigskip

KEYWORDS: Relativistic thermodynamics, generalized second law, Weyl curvature conjecture, gravitational entropy.

\section{Introduction}

Gravitational entropy is a measure of the degrees of freedom of the free gravitational field. In this sense, the notion of gravitational entropy is different from that of conventional matter entropy. In any gravitational process, for example, cosmological structure formation, the behaviour of the gravitational field in a local region of spacetime can be analyzed in terms of this quantity. An entropy of this kind is necessary for several reasons. Historically, this proposal was invoked to provide a proper sense of sequence to gravitational processes, such that the initial Universe may be in a state of low entropy, and subsequently the entropy would increase with the evolution of the Universe. Unless entropy is associated with the free gravitational field at the time of the formation of the Universe, a gravity dominated evolution of the Universe would violate the second law of thermodynamics (SLT).

The SLT is one of the most fundamental laws of physics and is also important from the point of view of the study of cosmological spacetimes. In the thermodynamic description of gravity, the Universe is expected to exist in a state of low entropy during the initial singularity (the Big Bang), and the  entropy should increase with the process of structure formation. According to big bang cosmology, the Cosmic Microwave Background Radiation (CMBR) is the relic of the ``big bang explosion'', and this background is very much isotropic in nature. In fact, the Universe was largely in a state of thermal equilibrium even after 300000 years of the big bang, and therefore, in a very homogeneous state. Gradually small density perturbations appeared due to the effects of gravity, thereby leading to structure formation and causing great temperature differences between the matter in the stars and in the interstellar medium. According to SLT, the entropy of the Universe should always increase. But the early Universe was very hot, with matter and radiation in thermal equilibrium, which signifies a state of maximum entropy. Does it mean that the universe violates the second law? The solution to this problem was provided by Roger Penrose who suggested that entropy must also be carried by free gravitational fields themselves, known as ``Gravitational entropy'' (GE), so that the SLT may be preserved. This proposal is known as the \textbf{\emph{Weyl Curvature Hypothesis}} (WCH) because according to Penrose the measure of GE should involve an integral of a quantity derived from the Weyl curvature tensor. Thus, the inclusion of GE not only solved the \emph{\textbf{initial entropy problem}} (the low entropy of the Universe in its initial state), but in addition to this, the necessity of incorporating the tendency of gravity to produce density inhomogeneities into the framework of a generalized second law of thermodynamics (GSLT) led Penrose to propose the Weyl curvature tensor as the measure of GE. The GSLT is basically the weakened version of SLT in which the sum of matter entropy and gravitational entropy has to remain non-decreasing during the evolution of the Universe.

The Weyl tensor is known to be vanishing for a homogeneous spacetime model, and is nonzero for an inhomogeneous universe. If the Weyl tensor is related to the GE, then the Universe would naturally evolve from a low entropy state in its initial homogeneous configuration to a state of higher entropy during the later stages when inhomogeneities set in due to structure formation. Any increase in GE would imply an increase in the local anisotropy of the Universe, represented by the Weyl tensor. The non activation of gravitational degrees of freedom at the big bang is accounted for in the WCH, according to which the initial spacetime singularities must be constrained to have a vanishing Weyl curvature (or some appropriate measure related to the Weyl curvature), whereas the final spacetime singularities (like those occurring inside black holes) are unconstrained. Therefore, after the big bang, the evolution of the Universe should be close to that of the homogeneous and isotropic Friedman-Lemaitre-Robertson-Walker (FLRW) model in conformity with the observed isotropy of the CMBR, and so be consistent with the SLT. However, there are several inflationary universe models with varying types of initial conditions that lead to conditions of approximate homogeneity after the end of inflation. This means that some suitable dimensionless scalar (which is related to the Weyl tensor) must be asymptotically zero. Therefore, the determination of the gravitational entropy function requires the construction of this scalar function.

Another importance of GE is that black hole entropy emerges as a special case of it. In this context it may be mentioned that black hole (BH) entropy belongs to a very distinctive class because it is the only entropy which is proportional to the surface area of a gravitating object unlike other thermodynamic entropies which are proportional to the volume of a physical system.

It turns out that a fully satisfactory and self-consistent definition of gravitational entropy in the context of cosmological structure formation is yet to emerge. Over the years, several candidates of GE have been proposed, initially based on Penrose's Weyl Curvature Hypothesis. But all such proposals have been applied to very restrictive and symmetric spacetimes. A few years ago, Clifton, Ellis, and Tavakol (CET) adopted a novel method of defining this entropy from the Gibbs equation written in terms of an effective stress-energy tensor that arises from the `square root' of the fourth order Bel-Robinson tensor, which is the simplest divergence-free tensor related to the Weyl curvature tensor. However, even this proposal is applicable only to restricted classes of spacetimes. Thus the problem of gravitational entropy remains as one of the open problems in relativistic gravity even today.

In this article we will trace the development of the concept of gravitational entropy, and review some of the notable works in this field, within the framework of General Relativity (GR).

\section{The Correlation between Gravity and Thermodynamics}
To discuss the emergence of the concept of gravitational entropy we begin from the initial efforts that went into the establishment of a relation between gravity and thermodynamics.

\subsection{Initial developments: The works of R. C. Tolman}
The correlation between gravity and thermodynamics was initiated primarily by Richard Chace Tolman, who wrote a series of papers to formulate the corresponding theory and wrote a book on this theory \cite{Tol1}.

In his paper ``On The Extension of Thermodynamics to General Relativity'', Tolman \cite{Tol2} mentioned that Wilhem Lenz \cite{Lenz} had published an interesting article on the equilibrium between radiation and matter in Einstein's closed universe. Lenz's article served as the motivation for Tolman to ``investigate the proper method of extending the ordinary principles of thermodynamics so as to make them hold for considerations in curved space-time where the methods of general relativity must be employed'' \cite{Tol2}. It appeared to Tolman that Lenz had relied on the principles of ordinary thermodynamics valid in flat space-time \cite{1}. So he assumed the state of equilibrium in the universe to be a state of maximum entropy for variations which did not affect the total energy. Consequently, the expressions he used for the energy and the entropy of the universe were similar to those in flat space-time, with the only difference that the ordinary volume of flat space was replaced by the corresponding volume of curved three-dimensional space.

However, it was well-known from the formulations in curved space-time that it is difficult to find a suitable expression for energy. Further, a proper expression for entropy conformable to Einstein's general relativity (GR) was also not available at that time. Therefore, in the words of Tolman, Lenz's proposal was inadequate for handling thermodynamic problems in curved space-time. This led Tolman to propose two principles expressed in terms of tensor equations (thus valid for all coordinate systems) which would be analogous to the ordinary first and second laws of thermodynamics in GR. So, let us review these two principles as proposed by Tolman \cite{Tol2}.

\subsubsection{The first law of relativistic thermodynamics}

This law would be the analogue of the laws of conservation of energy and momentum in flat space-time. In tensor notation, this law is expressed as:
\begin{equation}\label{01}
(T_{\nu}^{\mu} )_{, \mu}=0,
\end{equation}
where $T_{\nu}^{\mu}$ is the total stress-energy tensor for matter and radiation. To use equation \eqref{01} for physical systems, the vanishing divergence of the stress-energy tensor is converted into tensor densities and integrated over $4$-volume, following the methods described in \cite{Eddington}. In terms of tensor density $\mathcal{T}^{\nu}_{\mu\nu}$, \eqref{01} can be written in the form
\begin{equation}\label{01a}
\mathcal{T}^{\nu}_{\mu\nu} = (T_{\mu}^{\nu} )_{, \nu}\sqrt{-g} = \frac{\partial \mathcal{T}^{\nu}_{\mu}}{\partial x^{\nu}} - \frac{1}{2} \mathcal{T}^{\alpha\beta} \frac{\partial g_{\alpha\beta}}{\partial x^{\nu}} =0.
\end{equation}
Integrating over a specific four-dimensional volume, one obtains
\begin{equation}\label{01b}
\int \int \int \int \left ( \frac{\partial \mathcal{T}^{\nu}_{\mu}}{\partial x^{\nu}} - \frac{1}{2} \mathcal{T}^{\alpha\beta} \frac{\partial g_{\alpha\beta}}{\partial x^{\nu}} \right ) dx_1 dx_2 dx_3 dx_4 = 0,
\end{equation}
which is an invariant equation that serves as the analogue of the ordinary first law of thermodynamics as applicable to GR, and from which any investigation of the energy (and momentum) of a physical system may be initiated.

\subsubsection{The second law of relativistic thermodynamics: Tolman's definition of the entropy vector}

After finding an acceptable form of the first law of relativistic thermodynamics, Tolman sought to determine a proper expression for the second law of relativistic thermodynamics. However, soon he realized that this could not be done in the same way as in the case of the first law. Nevertheless, he still had two stringent criteria in hand which could be used as the guiding principles. First of all, the postulate had to be expressed in a covariant form. Secondly, it must be equivalent to the ordinary second law of thermodynamics in flat space-time. To ensure that the postulate of the generalized second law may satisfy these criteria, he defined the entropy vector at any given point in space-time by the following relation:
\begin{equation}\label{02}
S^{\mu}=\phi_0 \frac{dx^{\mu}}{ds},
\end{equation}
where  $dx^{\mu}/ds$  denotes the macroscopic motion of matter (or energy) at the given point, and $\phi_0$ is the proper entropy density as measured by a comoving observer. The divergence of the vector $S^{\mu}$ integrated over the $4$-volume of an isolated system (which includes the universe as a whole) should be positive or zero, subject to the condition that the conservation of the energy-momentum tensor remains valid.

The tensor density $\mathcal{S}^{\mu}_{\mu}$ corresponding to the divergence of $S^{\mu}$ is given by
\begin{equation}\label{02a}
(S^{\mu})_{, \mu} \sqrt{-g} = \mathcal{S}^{\mu}_{\mu} = \frac{\partial \mathcal{S}^{\mu}}{\partial x^{\mu}} ,
\end{equation}
and the integral of \eqref{02a} over a 4-dimensional region enclosing an entire isolated thermodynamic system, leads to the following mathematical form of the second law:
\begin{equation}\label{02b}
\int \int \int \int \frac{\partial \mathcal{S}^{\mu}}{\partial x^{\mu}} dx_1 dx_2 dx_3 dx_4 \geq 0 ,
\end{equation}
provided \eqref{01b} holds.

On careful examination, the above postulated equations were found to satisfy the criterion of being covariant in nature. Further, they reduced to the principles of ordinary thermodynamics in flat space-time, when these equations were considered in the Galilean coordinates.

\subsubsection{Tolman's subsequent works}

In the same year, Tolman published three more papers on related topics, one on the energy and entropy of Einstein's closed universe \cite{Tol3}, the second one on the equilibrium between radiation and matter in Einstein's closed universe \cite{Tol4}, and followed it up by adding further remarks to his formulation of the second law of thermodynamics in GR in the third article \cite{Tol5}. In 1930 he published four papers in the Physical Review journal, where he wrote on the use of the energy-momentum principle in GR \cite{Tol6}, on the use of the entropy principle in GR \cite{Tol7}, on the importance of the first law of thermodynamics in GR (a result which is popularly known as the ``Tolman law'') \cite{Tol8}, and along with P. Ehrenfest he wrote a paper on `Temperature Equilibrium in a Static Gravitational Field' \cite{TolEhren}. In 1931, he discussed in details the conditions of thermodynamic equilibrium in a static Einstein universe \cite{Tol9} and also the problem of the entropy of the Universe as a whole \cite{Tol10}. He also wrote an article on the applications of relativistic thermodynamics to cosmological problems \cite{Tol11}. He undertook further examination \cite{Tol11a} of the bearings of relativistic thermodynamics on the problem of the entropy of the Universe as a whole, and demonstrated that the framework of GR provides conceivable models of the Universe which undergo continued expansions and contractions without being brought to rest by irreversible processes that accompany these changes. This result was in contrast to the familiar result in classical thermodynamics which predicted that continued occurrences of irreversible processes would lead to a final state of maximum entropy and minimum free energy after which there will be no further change.

On December 29, 1932, he delivered the Tenth Josiah Willard Gibbs Lecture, at Atlantic City, under the auspices of the American Mathematical Society (AMS), at a joint meeting of the Society with the American Physical Society, and Section A of the American Association for the Advancement of Science, where he carefully explained the correlation between Thermodynamics and Relativity \cite{Tol11b}. With H. P. Robertson \cite{TolRob} he investigated the interpretation of the relativistic second law of thermodynamics given by
\begin{equation}\label{02c}
\int \int \int \int \frac{\partial \mathcal{S}^{\mu}}{\partial x^{\mu}} dx_1 dx_2 dx_3 dx_4 \geq \frac{dQ_0}{T_0} ,
\end{equation}
and showed that the quantity $dQ_0$ had the significance of the heat flowing relative to the fluid having instantaneous proper volume $dV_0$ during the proper time $dT_0$, such that $dV_0 dT_0 = \sqrt{-g} dx_1 dx_2 dx_3 dx_4$, as measured by a local observer at rest in the fluid at a chosen point, with $T_0$ as the temperature ascribed to this heat by the local observer. In a subsequent article he discussed the validity of the first law of thermodynamics in a general gravitational field \cite{Tol12}.

Tolman also worked on several solutions of the Einstein field equations and on cosmology \cite{Tolman_C1}-\cite{Tolman_C15}, as well as in other areas of physics, his last documented work being the paper \cite{Tolman_C16}. It may appear that Tolman was only interested in theoretical cosmology, where he made important contributions soon after Einstein's formulation of GR, and in the correlation between thermodynamics and relativity. However, in reality, he was simultaneously interested in a wide variety of topics, like statistical mechanics, applications of thermodynamics to chemistry, and in many important theoretical and experimental works that led to the solution of problems in chemical kinetics, and published several papers on those works, which have not been mentioned here.

\subsection{Developments in black hole thermodynamics and gravitational dynamics}
General relativity describes the dynamics of spacetime \cite{2} by the use of spacetime metric, while thermodynamics describes macroscopic properties of thermal systems in terms of parameters like energy, pressure, entropy, temperature, etc. Concrete evidence for the deep connection between gravitational dynamics and thermodynamics emerged from the theory of black hole thermodynamics (BHT), as illustrated by Bekenstein, Hawking, and others \cite{Bekenstein1,Bekenstein2,Hawking0,Hawking,BCH}.
According to BHT, a black hole (BH) has a temperature proportional to its surface gravity, and entropy proportional to the surface area of its horizon, and the temperature, the entropy, and the mass of the BH satisfies the first law of thermodynamics. The other three laws of thermodynamics are also found to be satisfied by BHs \cite{BCH}. However, a BH is a special thermodynamic system because its entropy is proportional to its horizon area, while the entropy of an ordinary thermal system is proportional to its volume. Another special feature of BHT is that the heat capacity of some BHs may be negative \cite{Hawking1,Lynden-Bell}, for example, the case of Schwarzschild black hole.

The realization that the Schwarzschild solution and its generalizations (those which have horizons) resemble thermodynamic systems \cite{Paddy1}, and the discovery of the evaporation of BHs that indicated a connection between the horizon, entropy and the temperature of such physical systems \cite{Hawking}, led to the development of the thermodynamics of spacetimes. The geometric feature of the temperature and entropy of a BH leads one to conjecture that gravity might be an emergent phenomenon, and is a coarse graining description of some microscopic degrees of freedom of spacetime \cite{Paddy,Hu}. This idea was first proposed by Sakharov in 1967 \cite{Sakharov}, who hinted that the background of spacetime geometry emerges as a mean field approximation of underlying microscopic degrees of freedom of spacetime. According to him, relativistic gravity (general relativity) emerges from quantum field theory in the same way that the theories of hydrodynamics and continuum elasticity emerges from molecular physics \cite{Visser}. In GR, the action of spacetime depends on the curvature. Sakharov said that this action leads to a ``metrical elasticity'' of spacetime, i.e., to generalized forces which oppose the curving of spacetime. Thus the action of GR was identified as the change in the action of quantum fluctuations of the vacuum, assuming spacetime to be curved. Sakharov considered the dependence of the action of the quantum fluctuations on the curvature of spacetime. This approach to gravity was analogous to the description of quantum electrodynamics in the papers by Soviet physicists E. S. Fradkin \cite{Frad1,Frad2}, L. D. Landau and I. Ya. Pomeranchuk \cite{LandPomer}.

The introduction of the Bekenstein-Hawking entropy \cite{Bekenstein1,Hawking} on the black hole event horizon led to the complete development of the thermodynamic picture of GR. Particularly in the case of BHs, it leads us to the well-known information paradox, because BHs with the same observational parameters may have evolved from different initial configurations along different paths of evolution and may be constituted of different internal parameters which are concealed inside the event horizon, so that some crucial information is lost.

The most striking similarity between black hole physics and thermodynamics lies in the behaviour of the BH horizon area and entropy, because both quantities tend to increase with time. The evolution of a BH usually takes place in the direction of increasing horizon area. If the universe is assumed to be a thermodynamic system bounded by a horizon, then a similar prescription should also be applicable to cosmological spacetimes. This idea facilitated the development of the thermodynamic treatment of cosmological evolution.

\section{ENTROPY FUNCTIONS IN RELATIVITY}

For a proper understanding of the sequence of development of the concept of gravitational entropy, we now try to gather an overview of the various entropy functions prevalent in relativistic gravity, most of which is already known. These are:
\begin{itemize}
  \item black hole entropy,
  \item entropy of cosmological horizons,
  \item gravitational entropy associated with the inhomogeneity of gravitational fields as a consequence of the Weyl curvature conjecture.
\end{itemize}
According to the generalized second law of thermodynamics (GSLT), the sum of all these entropies, together with the usual thermal entropy of the universe, should always increase.

\subsection{Blackhole entropy (Bekenstein-Hawking entropy)}

The topic of black hole entropy has been extensively studied by many experts. It is known that the Bekenstein-Hawking entropy is the entropy of a BH which is necessary for it to satisfy the second law of thermodynamics (SLT) as viewed by observers outside the BH. If black holes did not possess entropy, then the accretion of matter by a BH would lead us to the violation of the SLT. Treating black hole physics from the point of view of information theory, the concept of BH entropy was introduced as a measure of the information contained inside the BH, which is inaccessible to an external observer.

In 1972, J.D. Bekenstein \cite{Bekenstein1} used physical arguments to justify that an entropy function must be assigned to a black hole in order to preserve the validity of the second law of thermodynamics (SLT) on the BH horizon. The concept of BH entropy emerges from the generalized version of SLT, according to which if common entropy (entropy outside a BH) is engulfed inside a BH, the sum of the common entropy and the black hole entropy does not decrease. If the black hole possesses entropy, then its entropy should increase as it engulfs inflowing matter, while the entropy of the matter outside the BH decreases by the same amount. Thus BH entropy should have a genuine contribution to the total entropy of the universe. This generalized version of the second law is validated both from information theory, as well as from other examples like a harmonic oscillator composed of two particles connected by a nearly massless spring, or a beam of light directed towards a Kerr black hole, as illustrated by Bekenstein. He conjectured that BH entropy is proportional to the area of its event horizon divided by the square of the Planck length. Subsequently, in 1973, Bekenstein \cite{Bekenstein2} suggested that this proportionality constant has the magnitude of $ \ln 2 /8 \pi$. In the same year Bardeen, Carter and Hawking derived the laws of black hole mechanics \cite{BCH}. They found that two of the quantities appearing in the expressions, that is, the area $A$ of the event horizon and the surface gravity $\kappa$ of the black hole, are analogous to the conventional entropy and temperature of a physical system. From this analogy and subsequent justifications, they suggested that the four laws of black hole mechanics are \emph{\textbf{analogous}} to the four laws of thermodynamics.

Comparing the first law of black hole mechanics (Bardeen et. al \cite{BCH}), as applicable to two neighbouring BH equilibrium states, with the first law of thermodynamics, Stephen Hawking observed that if some multiple of $A$ is considered analogous to entropy, then some multiple of $\kappa$ is analogous to temperature. Moreover, $\kappa$ is analogous to temperature on one more count because it remains constant over the event horizon under equilibrium conditions. Arguing in favour of Bekenstein's conjecture \cite{Bekenstein2}, Hawking came to the conclusion that the proportionality constant should have the exact value of one-fourth \cite{Hawking}. He showed that quantum mechanical effects cause BHs to create and emit particles like heated objects at a rate that would be expected if the BH was assumed to be at a temperature of $(\kappa /2\pi)(\hbar /2K_B),$ where $K_B$ is the Boltzmann constant. Eventually this emission leads to the disappearance of the BH. Although these quantum effects violate the classical law that the area of the event horizon of a BH cannot decrease, but it is permitted by GSLT, according to which the sum of the entropy of matter outside the BH and the black hole entropy never decreases. This area behaves very much like the entropy of a physical system, but at that time it was difficult to reconcile the connection between a geometric quantity like area with that of entropy, which is a thermodynamic quantity. One reason for such difficulty in reconciliation is that BH entropy is numerically much greater than the entropy of any ordinary thermodynamic system of the same mass \cite{Bekenstein2}.

According to the cosmic no-hair theorem \cite{RW}, all stationary BH solutions are characterized by only three observable parameters. These are: its mass, electric charge and angular momentum. Thus BH entropy should also depend only on these parameters. These three parameters appear in such a combination in the expression of BH entropy that it may be identified with the surface area of the BH. If $A$ is the area of the event horizon of a BH, then BH entropy $S_{BH}$, in dimensionless units, is given by
\begin{equation}\label{04}
S_{BH}=A/(4L_P^2 )=(c^3 A)/4G\hbar,
\end{equation}
where $L_P=\sqrt{G\hbar/c^3}=1.6\times 10^{-35}~ m$ is the Planck length, $c$ is the speed of light in vacuum, $G$ is the Newtonian gravitational constant, and $\hbar$ is the reduced Planck constant, $h/2\pi$. For the stationary, spherically symmetric Schwarzschild BH, the only relevant parameter is its mass $M$, with the radius of the event horizon given by $r_h=  2GM/c^2 = R_s$,  where $R_s$ is the Schwarzschild radius of the black hole, and $\kappa = c^2/2R_s$ is the surface gravity at the horizon. The area of this horizon is given by
\begin{equation}\label{05}
A = 4 \pi r_h^2 = 16 \pi (GM/c^2 )^2.	
\end{equation}
In proper dimensional units, we have $$S_{BH}=\frac{\pi K_B R_s^2}{L_P^2} = \frac{\pi K_B \, c^4}{4L_P^2 } \left(\frac{1}{\kappa^2} \right).$$

Thus the Bekenstein-Hawking entropy of a BH is the product of the Boltzmann constant $K_B$ with one-fourth of the spatial area of its event horizon in Planck units. The horizon radius of one-solar mass Schwarzschild BH is $295\times10^3 m$ and hence its Bekenstein-Hawking entropy is $1.5\times10^{54} J / K$. This is about 20 orders of magnitude greater than the thermodynamic entropy of the sun, which therefore indicates that the formation of a BH is a highly irreversible process. Thus, BH entropy cannot be equal to the entropy of the physical system before it evolved to the BH state.

Bekenstein \cite{Bekenstein4} argued that it is difficult to imagine that two quantities that are so different in magnitude may be clubbed within the same bracket. Further, he added that even if the two entropies have similar origin, it is difficult to understand how can BH entropy be expressed statistically as the logarithm of the number of interior states, without taking the full account of everything that is possibly happening inside the BH. Pondering over whether BH entropy can be calculated in the same way as one calculates matter entropy, Bekenstein said that the apparent mismatch arises because BHs are being treated on the same footing as other nonrelativistic systems. In reality, BH entropy is the maximal limit for the entropy of an ordinary system of the same mass. This happens due to the existence of an upper bound to the entropy-to-energy ratio of `non blackhole systems' of given effective radius $R$. However, Unruh and Wald \cite{UW} pointed out that the Bekenstein bound is not necessary for the validity of the GSLT in the case of matter confined to a given volume of the Universe. They argued that in the context of classical BH physics, it is possible to immerse a box containing rest energy $E$ and entropy $S$ slowly into a BH, so that the box experiences an effective bouyancy force due to the acceleration radiation near the BH. Thus there will be a finite lower bound on the energy transferred to the BH and a minimal increase of the area of the BH, so that the GSLT is satisfied.

Hiscock \cite{Hiscock} considered the definition of the gravitational entropy (GE) of a nonstationary BH for a simple model in which a spherical shell collapses and enters a preexisting BH. Although the second law of BH mechanics \cite{BCH} identified one-quarter of the area of the event horizon as the GE of the BH, but in reality, it is impossible to locate the position of the global event horizon accurately from local measurements. Therefore, in the case of nonstationary BH, its GE was identified as one-quarter of the area of the apparent horizon, which is the boundary of the trapped region, or the outermost trapped surface, and the event horizon is the ``actual edge'' of the BH. The GE defined in this way will not decrease with time for realistic matter (for which the weak energy condition holds). In the case of stationary BH, the apparent horizon coincides with the event horizon. The difference between the `event-horizon-entropy' and the `apparent-horizon-entropy' could then be interpreted as the GE of the collapsing shell. Further, it could be verified that the GE of a solar-mass star would exceed its ordinary matter entropy about one year before it enters a galactic mass BH.

It was Wald \cite{Wald1}, who demonstrated that if GR is based on an action principle, then the first law of BH mechanics predicts that for perturbations of stationary BHs, the surface term at the horizon is always given by $\frac{\kappa}{2\pi}\delta S$, where $S$ is a local geometrical quantity of the horizon, and is equal to $2\pi$ times the Noether charge at the horizon of the horizon Killing field (normalized to yield unit surface gravity). The nature of $S$ indicated the possibility of a generalization of the definition of entropy to dynamical BHs. The relationship also indicated that the second law of BH mechanics in dynamical evolution from an initially stationary BH to a final stationary state, is equivalent to the positivity of the total Noether flux, and could be intimately connected with the positive energy properties of the theory.

In 1994 \cite{Hay0}, Hayward provided a general, dynamical, quasi-local definition of BHs in terms of ``trapping horizon'', and derived the general laws of black hole dynamics. The idea of trapping horizon is a concept similar to the apparent horizon. It is a hypersurface foliated by marginal surfaces of one of four non-degenerate types: future or past, and outer or inner. The future outer trapping horizon provides the definition of a BH. He gave a definition of the trapping gravity of the outer trapping horizon, thereby generalizing the concept of surface gravity. The first three laws of BH dynamics were formulated in terms of the behaviour of the trapping gravity and the area of the outer trapping horizon, aided with suitable energy conditions and topology. Thus in this new framework, BH dynamics was formulated as the dynamics of trapping horizons.

Hawking and Hunter \cite{HH} showed that entropy is a global quantity, like energy or angular momentum, and is not localized on the horizon. Attempts to identify the microstates responsible for BH entropy are nothing but constructions of dual theories residing in separate spacetimes. Further, the entropy arises from a failure to foliate topologically non-trivial Euclidean spacetimes with a family of time surfaces, in which case the Hamiltonian will not give a unitary evolution in time, thereby indicating the possibility of information loss and the issues of quantum coherence. In this situation there is an inequality between the Hamiltonian and the action per isometry period, the difference between them being the entropy.
Garfinkle and Mann \cite{GM} showed that this form of entropy (which they called ``generalized gravitational entropy'') could be interpreted as the integral of the Noether diffeomorphism charge integrated over all obstructions to the foliation.

\subsection{Entropy of the cosmological horizon}

The idea that gravitational field may have a role to play in the entropic evolution of the Universe came from the work on the evaporation of BHs by Stephen Hawking \cite{Hawking}, and Bekenstein's statistical interpretation of the concept of black hole entropy \cite{Bekenstein3}. This was followed by Penrose's suggestion \cite{Penrose00,Penrose3} that the SLT may be validated in terms of the behaviour of the gravitational field. Black holes are global structures that possess event horizons. Several cosmological models also have event horizons. Naturally one would ask whether the horizon area is still a measure of the entropy of cosmological spacetimes, and whether the GSLT holds for reasonable assumptions about the matter content of these spacetimes.

In 1977 Gibbons and Hawking \cite{GH} showed that the connection between event horizons and thermodynamics already observed in the case of BHs, can be extended to cosmological models, such that to an observer in these models the area of the event horizon (EH) will appear to be proportional to an entropy function, which may be identified with the absence of information available to the observer regarding the region beyond the EH. Moreover, the surface gravity $\kappa$ associated with the EH obeys a so-called ``first law'', which is similar to the manner in which the temperature of a thermodynamic system appears in the classical first law of thermodynamics. Subsequently, Bekenstein \cite{Bekenstein4} proved that an universal upper bound exists for the entropy-to-energy ratio of an arbitrary system of effective radius $R$.

The change in the area of the cosmological event horizon (CEH) was calculated by Davies \cite{Davies0} for cosmological models that are slightly different from the de Sitter space. For a perfect fluid obeying the dominant energy condition, the horizon area did not decrease. For a viscous fluid, although this area could decrease, but the entropy generation associated with this process ensured that the GSLT remained valid. He also proved an analogue of Hawking's area theorem for BHs in the case of FRW-type cosmological models with event horizons \cite{Davies00}. The GSLT was investigated for contracting models where the area was shrinking. It appeared that the horizon area could start decreasing long before the final singularity was reached. This might indicate a failure of the GSLT, which could be explained in two ways: (i) It might be that the horizon had no local significance except when its area settled down to a fixed or slowly varying value (e.g. the approach to de Sitter space). If the horizon was shrinking rapidly, the GSLT could not be saved by manipulating the entropy and area exchange. (ii) The failure of the GSLT might indicate either that the area of the event horizon was no longer a suitable measure of entropy, or it might still be so, but in that case it was necessary to augment it by an additional term, e.g. by attributing an entropy to account for the `degree of shrinkage' of the Universe, which increased to an unbounded value near the singularity, thereby extending the GSLT. Pollock and Singh \cite{PS} studied the thermodynamics of de Sitter and quasi-de Sitter spacetimes and found that the total entropy of de Sitter spacetime vanishes identically, as the positive entropy located beyond the event horizon was exactly complemented by the negative vacuum entropy contained within the event horizon. For the idealised eternally quasi-de Sitter models, the decrease in horizon entropy that occurred when the event horizon shrinked was compensated by an increase in the vacuum entropy. Their investigations revealed that there was a link between classical thermodynamics and the general theory of relativity, which was yet to be deciphered.

Thus, as a natural extension of the developments in BHT, it was possible to draw an effective analogy between black hole physics and the thermodynamics of the Universe modelled as a physical system bounded by a horizon. In fact spacetimes bounded by horizons behave in a way similar to thermodynamic systems, and one can assign a temperature and an entropy function to them. This led to the conjecture of the ``thermodynamics of horizons'' \cite{Jacobson,Paddy1}, according to which one can apply the laws of thermodynamics to cosmological horizons, assuming the Universe to be a system bounded by a causality barrier, instead of a diathermic barrier. The causality barrier implies that the system comprises of the degrees of freedom outside the horizon.
Therefore, the entropy $S_{CEH}$ of the cosmological event horizon having proper area $A_{CEH}$ should be given by
\begin{equation}\label{03}
S_{CEH}=\frac{K_B A_{CEH}}{4L_P^2}.
\end{equation}

Jacobson \cite{Jacobson} derived the Einstein equations from the proportionality between the BH entropy and the horizon area, assuming the thermodynamic relation $\delta Q = TdS$ to be valid for all local Rindler causal horizons through each spacetime point, where $\delta Q$ and $T$ are the energy flux and the Unruh temperature measured by an accelerated observer just inside the horizon. Along with Kang and Myers, he examined the zeroth law and the second law of BH thermodynamics for effective gravitational actions which included higher curvature interactions \cite{JKM}. They showed that for a BH which accretes positive energy matter in a quasistationary process, the entropy could never decrease irrespective of the details of the gravitational action.

The relation between gravitational entropy and the area of horizons was clarified by Hawking, Horowitz and Ross \cite{HHR}. They showed that the extreme Reissner-Nordstr\"{o}m BH possesses vanishing entropy, although its horizon area is nonzero. Further, the action which governs the rate of pair creation of extremal and nonextremal black holes, is directly related to the area of the acceleration horizon for the extremal case, and to the area of the black hole event horizon in the nonextremal case. Finally they argued that unitarity cannot be preserved in quantum gravity just by involving Planck scale curvature in the case of annihilation of black holes.

Hayward \cite{Hay} derived the unified first law of BH dynamics and relativistic thermodynamics in spherically symmetric spacetimes. Realizing that the first law of BH dynamics required a definition of surface gravity for non-stationary BHs, he introduced a new definition of surface gravity for spherically symmetric BHs, which had the same form as the definition of surface gravity in the case of stationary BHs involving the Killing vector on the Killing horizon, but instead used the Kodama vector on the trapping horizon for the non-stationary BHs.

A covariant entropy/area bound for all spacetimes satisfying Einstein equations with the dominant energy condition for matter, was constructed by Bousso \cite{Bousso1}. Locally the energy was not well-defined in GR, and it was necessary for the spacetime to possess an infinity for global definitions of mass, whereas it is always possible to define a covariant version of area in terms of the proper area of a surface. Therefore it was possible to construct an entropy/area bound instead of an entropy/mass bound. This bound could be saturated, but would not be exceeded in cosmological spacetimes and inside black holes. The basis of this construction was statistical, and it placed a limit on the number of degrees of freedom available in nature.

Padmanabhan \cite{Paddy2} formulated the first law on ``any'' horizon for a general static spherically symmetric spacetime, starting from the Einstein equations. This equivalence between the laws of thermodynamics and the laws of BH mechanics on one side and the Einstein equations on the other side, revealed a strong connection between quantum physics and gravity. He also developed a general formalism for addressing the thermodynamics of horizons in spherically symmetric spacetimes \cite{Paddy3}, and showed that the Friedmann equations in Einstein gravity can be derived from the Bekenstein-Hawking entropy \cite{Paddy4}.

Davies et al. \cite{DDL} tested the generalized second law (GSL) by investigating the change in entropy when dust, radiation and black holes cross a cosmological event horizon. Using numerical calculations they determined the evolution of the cosmological horizon for flat, open and closed FRW universes, and found that in most cases the loss of entropy from within the cosmological horizon is more than balanced by an increase in the entropy of the cosmological event horizon, thereby maintaining the validity of the GSLT.

This was followed by a general ansatz for GE provided by Padmanabhan \cite{Paddy} who considered that any section of area which serves as the horizon for an appropriately defined accelerated observer, must possess an entropy proportional to this area. The definition of the entropy associated with the horizon as perceived by a congruence of observers, was introduced. This quantity was related to the GE under a set of conditions that included the identification of the horizon as that of a BH, in which case the GE is related to the surface gravity of the BH. Similar prescriptions are applicable to the case of Rindler observers. The relation between the entropy $S$ and the energy source $E$ of the gravitational field for any static spacetime with an associated temperature $\beta^{-1}$, was derived. The standard results known for BHs could be generalized in this approach, and further it provided a link to the holographic interpretation of spacetime \cite{Bousso2}.

It is known that the cosmological event horizon does not exist in big bang cosmology, but in an accelerating universe dominated by dark energy with a
time-dependent equation of state, the event horizon separates out from the apparent horizon. In this context, assuming the physical universe to be bounded by the dynamical apparent horizon, Wang et al \cite{WGA} showed that both the first and the second laws of thermodynamics hold on the apparent horizon. However, for a universe bounded by the CEH, both these laws break down at the event horizon if the usual definition of temperature and entropy that is applicable to the apparent horizon, is extended to the event horizon.

Egan and Lineweaver \cite{EL} used the mass function of supermassive black holes (SMBHs) to determine their entropy contribution to the entropy of the observable universe for the FRW model, and found that SMBHs are the largest contributor. They considered two different schemes for applying the GSLT. In the first scheme they assumed that the total entropy in a sufficiently large comoving volume of the Universe does not decrease with cosmic time. The Universe is bounded by a closed comoving surface and is effectively isolated because there is no net inflow or outflow of entropy from the comoving volume due to the large scale homogeneity and isotropy of the Universe. The second scheme assumes that the total entropy of matter contained within the CEH plus the entropy of the CEH itself, does not decrease with cosmic time. In this case the Universe is assumed to be bounded by the time-dependent CEH instead of a comoving boundary. The authors determined the entropy budget of the universe under both schemes. The first budget was consistent with previous estimates available in literature with the exception that SMBHs dominated the budget, and contributed at
least an order of magnitude more entropy than what was estimated previously. The second budget was dominated by the CEH entropy, which was an additional component in the second scheme, which also corresponded to the holographic bound (proposed by ’t Hooft in 1993) on the possible entropy of the other components and might have represented a significant overestimate in the entropy of the Universe. Lineweaver and Egan \cite{LE} computed the entropy of the Universe including the entropy of the current event horizon and the entropy of the matter and photons within the CEH, and found that the entropy of the current CEH is $\sim 10^{19}$ times larger than the next most dominant contribution, which is from the SMBHs. They plotted the entropy budget as a function of time and found that the CEH entropy had dominated other sources of entropy since $10^{-20}$ seconds after the big bang.

Considering a homogeneous and isotropic universe with the event horizon as the boundary, having perfect fluid matter content and an arbitrary equation of state, Mazumder and Chakraborty \cite{MC1,MC2} showed that the GSLT is valid for this universe if the first law of thermodynamics is satisfied by this system, but there will be some restrictions on the matter content. Mazumder et al. \cite{MBC} analyzed the thermodynamics of an inhomogeneous universe described by the spherically symmetric LTB model, and found that the trapping horizon coincides with the apparent horizon. They showed that the Einstein field equations are equivalent to the unified first law of thermodynamics. They examined the validity of the GSLT by assuming that the first law of thermodynamics holds at the apparent horizon for the perfect fluid and at the event horizon for holographic dark energy.

An important scenario of cosmological evolution known as the ``emergent universe'' was proposed by Ellis and collaborators \cite{EM,EMT}, in which the Universe originates from an Einstein static phase such that the scale factor of the FRW metric does not vanish, and as a result important physical quantities like the energy density and pressure do not diverge, unlike the hot big bang model. The paper \cite{dC_H_P} examined whether the GSLT was fulfilled in the transition from the initial Einstein static phase to the inflationary phase, and from the end of inflation to the era of thermal radiation dominated expansion. It was found that the condition for the validity of GSLT was that the radiation component of energy had a smaller contribution to the total energy density of the static phase.

Mathew and others \cite{MAV} studied the time evolution of the CEH entropy and the status of GSL for two types of flat Friedmann universes characterized by a positive cosmological constant, the first one consisted of radiation and the second one was composed of non-relativistic matter. They showed that the GSL constrained the temperature of both the radiation as well as the non-relativistic matter for the two different Friedmann universes. They also found that even though the net entropy of the radiation or matter was decreasing at sufficiently large times as the universe expanded, but it was increasing at early times when the universe was undergoing decelerated expansion. Thus the entropy of the radiation within the comoving volume underwent a decrease only when the universe experienced an accelerated expansion. The status of the GSLT was also analyzed for a cosmological model dominated by a stiff fluid and having constant bulk viscosity with the apparent horizon as the boundary of the universe \cite{MAM}, and it was found that the law was generally satisfied for suitable values of model parameters.

In 2015, Bousso and Engelhardt \cite{BE} proved a new area law in GR considering that any future holographic screen is a hypersurface of indefinite signature, foliated by marginally trapped surfaces. They proved that the area of these trapped surfaces increases strictly monotonically along the foliation. Future holographic screens arise in gravitational collapse of matter and near the future singularity called the Big Crunch. Past holographic screen exists in any expanding universe like the one in which we live. Both screens obey similar area law. Assuming the null curvature condition and certain generic conditions within the framework of GR, the authors proved several intermediate results and finally established that if the surfaces dividing the entire spacetime into a future interior and a past exterior are marginally trapped, then the evolution \textbf{must} be everywhere to the past or exterior, and the area theorem followed. This was the first rigorous area law that represented the GSLT in big bang cosmology. Unlike event horizons, these holographic screens are local as they can be identified at a finite time and without reference to an asymptotic boundary.

Following on this work, Carroll and Chatwin-Davies \cite{CC-D} established a connection between the tendency of cosmological spacetimes having a positive cosmological constant to evolve toward an asymptotically de Sitter phase, and the increase of entropy over time in such spacetimes (as  the de Sitter spacetime represents a maximum-entropy state). They proved that if the generalized entropy (in the cosmological version of the generalized second law proposed by Bousso and Engelhardt), increases to a finite maximum value along a quantum holographic screen that has specific entropic properties (as admitted by the FRW and Bianchi I spacetimes), then the spacetime is asymptotically de Sitter in the future. The limiting value of this
generalized entropy was found to coincide with the de Sitter horizon entropy.

\subsubsection{The initial entropy problem}

Although a suitable definition of GE exists for stationary black holes, but a universally acceptable analogue in the case of cosmology is yet to be defined. The initial state of the Universe was very special, and any proposal for the actual nature of this initial state must account for its extreme special condition. This state was extremely homogeneous, and later on, small density fluctuations appeared due to the effect of gravity, leading to structure formation. This evolution contradicts our expectations from the thermodynamic point of view, since gravitating matter gets condensed, unlike an ideal gas confined to a closed chamber, which spreads out after the chamber is opened, reaching a state of maximum entropy.

In the language of Roger Penrose \cite{Penrose1}, if $B$ is the phase space for the initial big bang singularity and $P$ is the total phase space of the universe as a closed system, then the volume of $P$ is greater than that of $B$ by a factor that exceeds $\left.10^{10}\right.^{123}$. This indicates the enormity of precision in the big bang. However, the best evidence for the possible occurrence of the big bang comes from the observations of the cosmic microwave background radiation (CMBR). The intensity of this radiation, when expressed in terms of frequency, agrees with the Planck radiation formula very closely, indicating that the early universe was composed of matter in thermal equilibrium. Such a state corresponds to maximum entropy. This appears to be inconsistent with SLT, according to which the Universe should have begun with a very small value of entropy.

The large scale homogeneity of the Universe implies a negligible flow of entropy between volumes greater than a scale of several hundred Megaparsecs. It is assumed that this large representative volume is closed, so that the SLT is valid on the cosmological horizon (in particular, the apparent horizon) \cite{WGA,ChakGuha}. To be specific, the entropy of the observable universe $S_{univ}$ (defined as the entropy of the comoving volume of our current particle horizon) should not decrease \cite{Penrose2}. Thus the early universe must have been in a state of low entropy \cite{Luminet,Vaas,Lineweaver}. The increasing entropy of the Universe would eventually approach a maximum value given by $S_{max}$ \cite{Bekenstein2}. This entropy change being unidirectional, naturally points towards the notion of an arrow of time.

The photons in the CMBR have temperature fluctuations of about $10^{-5}$ around their average blackbody temperature of 2.7 K \cite{Smoot,Fixen}. The entropy of a given comoving volume of blackbody photons in an expanding universe remains constant and represents the maximum entropy for the photons \cite{KolTur}. Thus, the CMBR spectrum is remarkably close to an equilibrium blackbody spectrum even about 400,000 years after the big bang, thereby indicating a universe close to thermal and chemical equilibrium with ``apparently'' high entropy during its initial stages. This observation is contrary to our expectations and is known as the \emph{\textbf{initial entropy problem}} (IEP). According to the ``Past Hypothesis'' \cite{Penrose3,Earman,Ainsworth}, the Universe originated in a special state of low entropy, in agreement with the SLT. Actual calculations indicate \cite{Wallace,AG} that the entropy of the universe increases during the process of structure formation, on account of gravitational collapse \cite{Davies}.

The IEP may be addressed to in the following way \cite{Riggs}: During the Planck era, shortly after the big bang, the size of the causal region of the Universe was of the order of the Planck length. The entropy of the observable universe was at a maximum value at that time due to the streaming radiation. However, the Universe at that time was devoid of clumped structures like stars and galaxies, and therefore the number of accessible states of the Universe were very limited. This feature justifies the low value of entropy of the Universe at that time compared to today's value. Thus the solution to the cosmic IEP should be able to explain both the initial low entropy of the Universe (as is required by the SLT), and the apparent high entropy of the CMBR. The observations of the CMBR indicate a universe at thermal and chemical equilibrium at the Planck era, which demands maximum entropy. This problem may be circumvented by including the low gravitational entropy of the homogeneous distribution of matter in the early universe and define a new maximum of entropy that includes the \textbf{\emph{gravitational entropy}}. So we have \cite{PatLine}:
\begin{equation}\label{06}
S_{max}=S_{(max,GRAV)}+S_{(max,CMB)}.
\end{equation}
Further, we also require that $S_{(max,GRAV)}\gg S_{(max,CMB)}$.

The high entropy of the CMBR cannot refer to the entropy of the gravitational field because gravity was negligible at the time shortly after the big bang. The early universe was of extremely uniform distribution because the gravitational degrees of freedom of the universe were not yet excited. As time passed, these degrees of freedom got slowly excited so that matter began to condense, leading to structure formation and for a sufficiently large system, its corresponding entropy increased. The binding energy released in this process heated up the system. Stars were formed, which at the end of their life cycle collapsed to black holes, which represent the states of maximum entropy.

Substantial reduction of GE may also have taken place during the time of inflation, although the thermodynamic entropy increased. At the end of inflation the universe attained a state of thermal equilibrium and sufficient homogeneity. After that, further condensation of matter due to gravitational effects led to the increase in local inhomogeneities and the release of gravitational energy, as a result of which the thermal equilibrium got disturbed and the thermodynamic entropy also increased. The gravitational entropy increased at the same time due to gravitational condensation, and eventually resulted in the formation of so many black holes in the Universe.

\subsubsection{The arrow(s) of time}

The development of thermodynamics following the works of Ludwig Boltzmann (see \cite{Boltzmann,Planck} for details) led physicists and philosophers to delve deep into the concept of the arrow of time. In his essay, Gold \cite{Gold} discussed the time symmetry of physical laws, explaining that any asymmetry is predominantly confined to the arena of statistical mechanics on account of the tendency of radiation to spread out, which in turn is related to the expansion of the Universe, there being no visible asymmetry in the laws of electrodynamics. He cited a number of examples to illustrate his point. According to him, the entropy would decrease if the Universe entered a phase of contraction in the future, resembling a bouncing scenario. To this extent, he overlooked the asymmetry observed in various aspects of nature.

Several arguments on the issue of the arrow of time have appeared in the past (see \cite{Vaas,Price1,Price2} and references therein for related discussions). Davies \cite{Davies1} offered an explanation for the origin of the observed time asymmetry in the Universe on the basis of the inflationary scenario of the early universe \cite{Guth}. This explanation was disputed by Don Page \cite{Page1}, who noted that the inftationary scenario itself invokes time asymmetry due to the assumption of the absence of initial spatial correlations, as a result of which the charge-parity-time (CPT)-invariant dynamical laws cannot explain the time asymmetry other than postulating these special initial conditions, as demanded by Penrose. He observed that \cite{Page2} there were also difficulties in explaining the second law on the basis of inflation, which describes the time asymmetry of the Universe, because inflation must assume time asymmetry rather than explaining it. However, Davies maintained that if the special initial conditions are compatible with the requirements of a quantum gravity regime, and are succeeded by an inflationary phase, then that would lead the Universe to a situation from which the present condition of increasing entropy becomes a natural consequence, thereby justifying the time asymmetry.

Although there are several possible arrows of time (Vaas \cite{Vaas} has listed at least ten such), the problem of gravitational entropy is concerned with the thermodynamic arrow of time. However, the thermodynamic arrow is related to the cosmological arrow of time \cite{Hawking2}, as argued by Hawking who showed that the arrow of time defined by the direction in which entropy increases is related to the cosmological arrow of time defined by the direction of time in which the Universe is expanding. Further, it was also shown that density perturbations give rise to a thermodynamic arrow of time that points in a constant direction while the Universe expands and contracts again \cite{HLL}. The philosophical explanation of such an occurrence could only be invoked from the weak anthropic principle \cite{Carter}.

Carroll and Chen \cite{CarrChen1} suggested that spontaneous eternal inflation could provide a natural explanation for the thermodynamic arrow of time. They discussed at length the underlying assumptions and consequences of this view. However, they were also conscious that inflation by itself cannot explain the time asymmetry of the Universe as it required the initial state to be a special one of low entropy \cite{CarrChen2}, instead of being an arbitrarily chosen state. According to them, a favourable situation is one in which inflation arises via thermal fluctuation in a background de Sitter spacetime, so that the Universe arises as a quantum fluctuation from an initial vacuum state. The increase of entropy leads to the creation of new patches, which eventually evolve into universes similar to ours, and the universe appears time-asymmetric to us because we can only see a small portion of it. In his paper \cite{Wald2}, Wald dealt with the condition of a special initial state of our universe, including the issues of the existence of the thermodynamic arrow of time, the low entropy of the present universe and even lower entropy of the early universe.

Bouncing scenario was considered by Hartle and Hertog \cite{HarHer}, where the entropy can decrease to a point, violating the SLT, and the arrow of time defined by the growth in the fluctuations of the wave function of the Universe is bidirectional. Vilenkin \cite{Vilenkin} considered two cosmological scenarios with bidirectional arrows of time which underwent a de Sitter-like bounce, with the thermodynamic arrow of time pointing in opposite directions away from the bounce. The first scenario (suggested by Aguirre and Gratton (AG) \cite{AG1,AG2}) assumed that there were two asymptotic regions separated by a de Sitter-like bounce, with low-entropy boundary conditions imposed at the bounce. The second scenario (proposed by Carroll and Chen), assumed generic initial conditions on an infinite spacelike Cauchy surface. Vilenkin assumed the null convergence condition to show that the Cauchy surface in a nonsingular universe (other than a BH) with two asymptotically inflating regions, would necessarily be compact, and the size of the universe at the bounce between the two asymptotic regions would be nearly the same as that of the de Sitter horizon. This would imply that the spacetime structure was similar to the AG scenario with no need of any special boundary conditions at the bounce.

Barbour et al. \cite{BKM1} presented the evidence for a conjecture that is alternative to the `past hypothesis', in that the arrow of time exists in all solutions of gravitational field equations governing the evolution of the Universe. They argued that such arrows arose due to asymmetry in the ``space of its true degrees of freedom'' (the shape space). They proved that their conjecture was valid for arrows of complexity and information in the Newtonian $N$-body problem, specifically in the vacuum Bianchi IX model. Such a proposition was novel in its approach and thorough in its analysis. The authors resolved the problem of a time-symmetric law that leads to observationally irreversible behaviour. They also indicated how the other arrows of time could arise. They dealt further on this issue in another paper \cite{BKM2} where they showed that it is not necessary to impose any special initial conditions on any time-symmetric law for its solutions to exhibit a behavior that defines an arrow of time. For this purpose they considered the Newtonian $N$-body problem with vanishing total energy and angular momentum, for which all solutions divide at a unique point into two halves. In each half, a well-defined measure of shape complexity fluctuates but grows irreversibly between rising bounds from that point. Each solution would have a single past and two distinct futures emerging from it. Any internal observer must be in one half of the solution and will only be aware of the records of one branch and deduce a unique past and future direction by inspecting the available records.

In their paper, Goldstein et al. \cite{GTZ} discussed the possible explanation for the origin of the thermodynamic arrow of time proposed by Carrol and Chen (CC) \cite{CarrChen1,CarrChen2}, and explored by Barbour et al. (BKM) \cite{BKM1,BKM2}. In the models proposed by CC and BKM, the Universe experienced an increase in entropy as required by the SLT, but there was no assumption on a special low entropy initial state of the Universe. The authors in \cite{GTZ} used mathematical, evidential, and theoretical reasoning to explain that in the models of CC and BKM, nearly all possible histories of the Universe were such that the entropy curve as a function of time would initially decrease from positive infinite limit to a minimum value and then diverge to positive infinite limit, giving rise to a thermodynamic arrow of time in each of the two temporal halves of evolution.

\subsection{Penrose's Weyl Curvature Hypothesis and the emergence of gravitational entropy proposals}
Although, there is no generally accepted measure of the entropy of the free gravitational field in general relativity, it is still possible to explain the \emph{\textbf{non-activation}} of gravitational degrees of freedom at the time of big bang. If the evolution of the universe is dominated solely by gravity, then we may encounter a violation of the SLT, if we are considering the thermodynamic entropy only.

The works on BH entropy and the entropy of the Universe had already shown us that it is possible to associate an entropy function with the unavailability of information about matter fields or even gravitational field itself. It was Penrose who conjectured that the Weyl curvature may serve as a measure of gravitational entropy. Through his arguments he clarified the relationship between gravitational condensation and the increase of entropy \cite{Penrose3,Penrose2a}. As gravity leads to clumping of matter, the measure of the degree of gravitational condensation should be a measure of GE. The absence of clumping implies spatial isotropy, which means that there are no gravitational principal null directions, and therefore corresponds to a vanishing Weyl curvature. When a BH is formed at the end state of gravitational collapse, the Weyl curvature diverges to infinity at the final singularity. As the time evolution of the Universe is unidirectional (time-asymmetric), then such a universe would follow time-symmetric physical laws if the initial conditions are time-asymmetric. Hence, only those cosmologies are physically acceptable for which the Weyl curvature vanishes at the initial singularity and increases thereafter. Thus the favoured model for the evolution of the Universe after the bang \cite{Tod1,ES} turns out to be the homogeneous and isotropic FLRW model, for which the Weyl tensor is zero at the big bang. Further, it appears that \cite{ES} the geometry must be exactly FLRW in the neighbourhood of the big bang, which means that we require additional mechanisms to account for the growth of inhomogeneity that would lead us to the present state of the Universe.

In order that the Weyl curvature may be identified with the GE, it must satisfy the criteria which are usually associated with thermodynamic entropy. First of all, it should increase monotonically with time from a finite or zero value at the initial singularity. Secondly, its value should increase with the energy of the gravitational field, and finally, a higher degree of gravitational condensation of matter should correspond to higher value of the Weyl curvature. In the absence of gravitation, a homogeneous state is a state of maximum entropy, whereas in the presence of gravity, physical systems have a natural tendency to evolve from a state of uniform distribution to states of greater inhomogeneity arising due to clumping of matter. To provide a proper sequence to the occurrence of gravitational processes, incorporating the tendency of attractive gravity to give rise to inhomogeneities (structure formation including the formation of stars and black holes), Penrose conjectured that the entropy function for the free gravitational field should vanish for a homogeneous distribution and attain a maximum value given by the Bekenstein-Hawking entropy in the case of a BH.

Therefore, according to Penrose's \emph{Weyl curvature hypothesis} \cite{Penrose3}, the Weyl curvature tensor $C_{\mu \nu \rho \sigma}$ (or some appropriate measure related to it, for example, the Weyl scalar $C_{\mu \nu \rho \sigma} C^{\mu \nu \rho \sigma}$), must vanish at the big bang singularity whereas it may be unconstrained at the final spacetime singularities (cosmological, like the big crunch singularity or astrophysical, like the black holes). The idea of non-activation of the gravitational degrees of freedom at the time of big bang is then consistent with the WCH. An increase in GE would imply an increase in the local anisotropy of spacetime, which can be quantified by the shear tensor. From the analysis of the trace free Bianchi identities we can infer that this shear tensor affects the evolution of the Weyl tensor \cite{MB}. This means that there must be a physical relationship between local inhomogeneities represented by the shear tensor, and gravitational entropy expressed in terms of the Weyl curvature tensor.

\section{PROPOSALS OF GRAVITATIONAL ENTROPY AND THEIR SIGNIFICANCE}
The proposals of gravitational entropy can be provided both in the local as well as in global contexts. Locally, the entropy of any astrophysical object (e.g. a star or a BH) indicates the enormous amount of entropy confined to a given region of spacetime due to gravitational condensation in that region, leading to the increase of entropy of the universe. BH entropy signifies the maximum value of this entropy for a given amount of matter in a given region of spacetime. Global studies on the evolution of the Universe also indicate that the GE is a monotonically increasing function and well behaved near the initial singularity.

The study of GE is also important from the thermodynamic point of view. We know that geometry and energy are interrelated in gravitational theories, and both geometric and thermodynamic measures are available for the GE. Through this study it is possible for us to estimate the overall energetics of a region of spacetime, and also understand the effect of the specific geometry, i.e., how matter and free gravitational fields behave either in a given region of spacetime or on a large scale in the universe.

The Weyl curvature hypothesis offered us the first hint to formalize the concept of GE. We know that the Universe began from a singular state, and the Weyl component of the Riemann tensor was much smaller compared to the Ricci component during the initial phase of the universe. This idea seemed quite tenable because the Weyl tensor is independent of the local energy–momentum tensor. The high isotropy of the CMB radiation indicate that the universe was in a nearly homogeneous state before structure formation began. Naturally one would expect that the evolution of the universe must be very close to that of a FLRW model, and be consistent with the SLT \cite{PL,PC}. Further, the Weyl curvature is zero at early times for the FLRW case but is large in the case of Schwarzschild-like spacetimes, which represents the geometry outside a spherically symmetric star or a BH formed in the later phase of evolution of a star. Thus the Weyl curvature hypothesis seemed to be justified, because the GE will be larger in strongly gravitating systems than in a flat spacetime, in agreement with the SLT \cite{Penrose4}.

Following Penrose's suggestion that a viable measure of GE should involve an integral of a quantity derived from the Weyl curvature tensor, several researchers have tried to construct integrals of scalar invariants related to the Weyl curvature tensor, which would serve as a suitable measure of GE. These proposals originated primarily from the necessity to preserve the second law of thermodynamics as applicable to the universe as a whole. In spite of all these efforts, there is still doubt about the definition of gravitational entropy in a way similar to thermodynamic entropy, which may be applicable to all gravitating systems, as pointed out by Clifton et al. \cite{CET}. In addition to this, as noted by Davies \cite{Davies1}, since attractive gravity leads to condensation of matter, therefore, a homogeneous distribution of matter/energy represents a configuration of lower GE. On the other hand, repulsive gravity tends to smoothen out the distribution, and in such a scenario, a state of regular distribution of matter/energy corresponds to higher value of GE. Thus the scenario in repulsive gravity is directly opposite to the notion of GE that follows from the WCH.

Therefore, it is evident that the proposals of gravitational entropy require further refinement and development.

\subsection{The Weyl scalar as the Gravitational Entropy}

The Weyl tensor is the independent component of the Riemann tensor that is not captured by the Ricci tensor. It is the traceless part of the Riemann curvature tensor, and in $n$ dimensions it is expressed as \cite{Chandra}
\begin{equation}\label{07}
C_{\alpha\beta\gamma\delta} = R_{\alpha\beta\gamma\delta} + (1/(n-2) )(g_{\alpha\gamma} R_{\beta\delta} + g_{\beta\delta} R_{\alpha\gamma} - g_{\beta\gamma} R_{\alpha\delta} - g_{\alpha\delta} R_{\beta\gamma}) + (1/(n-1)(n-2))  R(g_{\alpha\gamma} g_{\beta\delta} - g_{\alpha\delta} g_{\beta\gamma}),
\end{equation}
where $n$ is the number of spacetime dimensions, $R_{\alpha\beta\gamma\delta} $ is the covariant Riemann tensor, $R_{\alpha\beta}$ is the Ricci tensor and $R$ is the Ricciscalar.

Over the years, several definitions of gravitational entropy constructed from the Weyl curvature have been proposed, the simplest choice being that of the Weyl scalar:
\begin{equation}\label{08}
 W = C_{\mu \nu \rho \sigma} C^{\mu \nu \rho \sigma}.
\end{equation}

In 1988, Husain \cite{Husain} examined the WCH in the case of Gowdy space-times. He calculated the expectation values of the square of the Weyl curvature tensor for states of clumped and unclumped gravitons and found that the Weyl square contains information about the gravitational entropy, as it can be used to examine time asymmetry. But this choice has been criticized \cite{Rothman,RA}, as it was found that for a perturbed flat space metric (with gravitational wave perturbations) of the type
\begin{equation}\label{09}
ds^2  = a^2 (\eta)[-(1 + 2\Phi(\eta,z))d\eta^2+(1- 2\Phi(\eta,z)) \gamma_ij dx^i dx^j  ],
\end{equation}
with
\begin{equation}\label{09a}
\gamma_ij = \delta_ij \left[ 1 + \frac{\kappa}{4} (x^2 + y^2 + z^2 ) \right]^{-2},
\end{equation}
where $\eta$ is the conformal time, $a(\eta)$ is the expansion scale factor, $h_{ij} \ll \delta_{ij}$ represents the gravitational wave perturbations, $\Phi$ is the gauge-invariant version of $h$, and $\kappa = 0, -1, 1$ for flat, open and closed universes, respectively, the Weyl scalar is given by
\begin{equation}\label{10}
W=C_{\mu \nu \rho \sigma} C^{\mu \nu \rho \sigma} = (16(\Phi_{,zz} )^2)/(3a^4 ),
\end{equation}
which decreases with increasing time and inhomogeneity. The same result as in \eqref{10} was obtained by \cite{RA1} for density perturbations in dust-filled models in the longitudinal gauge for a flat expanding universe with perturbations only along the $z$ direction. This result does not agree with the WCH, as it does not produce the right behavior for the growing modes, for which the Weyl tensor decreases with increasing time and inhomogeneity. The problem could not be rectified by introducing an overall minus sign, as in that case the decaying modes increase with decreasing inhomogeneity, which is undesirable. Similar feature was also observed by \cite{RA1} for spatially flat but expanding gravitational wave spacetime.

\subsection{Gravitational Entropy in terms of a dimensionless scalar}

In 1984, Wainwright and Anderson \cite{WA} proposed a measure of GE in terms of the ratio of the Weyl square and Ricci square:
\begin{equation}\label{11}
P^2 = C_{\mu \nu \rho \sigma} C^{\mu \nu \rho \sigma}/R_{\mu\nu} R^{\mu\nu} .
\end{equation}
According to them, the quantity $P^2$ would vanish at the time of the Big Bang.
Goode and Wainwright \cite{GW} considered an exact solution of the Einstein field equations studied by Liang in 1972 \cite{Liang} with an irrotational fluid source described by the equation of state $p = \mu/3$, energy density  $\mu = T_{ab} u^a u^b = 3/(4t^2 A^2)$, and fluid velocity $u = A^{-1/2}  \partial/\partial t$, representing the radiation dominated universe, and whose initial singularity is `Friedmann-like'. They gave the geometric definition of the concept of `isotropic singularity' and showed that the Weyl tensor is dominated by the Ricci tensor at this scalar polynomial curvature singularity. They showed that at an isotropic singularity the magnetic part of the Weyl tensor vanishes and the electric part becomes identical to the trace-free three-dimensional Ricci tensor of the singularity. Hence the condition of zero Weyl tensor at the singularity rendered it to be a manifold of constant curvature.

From the above works it followed that the Weyl tensor (or some function of it) would have an important role to play in the evolution of self-gravitating systems (see for example refs. [1-4] in \cite{Herrera}, which includes \cite{WA,GW} along with other important references). Considering spherically symmetric collapse of matter cloud from an initial state of equilibrium (or quasi-equilibrium), Herrera \cite{Herrera} showed that the matter distribution would leave the state of equilibrium (or quasi-equilibrium) if and only if the Weyl tensor underwent a change with respect to its value in the initial state of equilibrium (or quasi-equilibrium). Herrera and co-authors \cite{HDiPMOST} provided the necessary, and sufficient condition for the vanishing of the spatial gradients of energy density during the evolution of self-gravitating spherically symmetric dissipative fluids with anisotropic stresses, with emphasis on the Weyl tensor, the shear tensor, the local anisotropy of the pressure, and the density inhomogeneity, which suggested a possible definition of a \emph{gravitational arrow of time}. In the case of charged, dissipative, spherically symmetric gravitational collapse with shear \cite{DipHLeDMacS}, they showed that the Weyl scalar depends on the charge within the collapsing cloud, so that any presence of electric charge would affect the Weyl tensor and contribute to the \emph{gravitational entropy}. For the non-dissipative locally isotropic fluid \cite{HODipFT}, the vanishing of the energy density inhomogeneity was found to correspond to a vanishing Weyl tensor, which could be related to the fact that tidal forces make a self-gravitating fluid more inhomogeneous as the collapse proceeds, so that the gravitational arrow of time indicating the progress of collapse could be defined in terms of the Weyl tensor.

To take into account the situation in realistic cosmological models containing anisotropy or inhomogeneity, for which the Weyl tensor might not necessarily vanish at the initial singularity, Bonnor \cite{Bonnor0} considered the recollapsing Tolman cosmological models \cite{Tolman_C10} in which he excluded the presence of white holes. He constructed an invariant quantity out of the Weyl and the Ricci tensors which turned out to be a non-decreasing function of cosmic time. He also defined a vector that could describe the gravitational entropy flux in these models. For such models, the gravitational epoch function $P$ would be a non-decreasing function of cosmic time so as to agree with the WCH, and vanish at the big bang, provided that the initial state of the universe represented an isotropic singularity \cite{GW}. This in turn meant that the universe began from a highly regular initial state, evolving steadily away from regularity on account of gravitational instability. In the scenario of `quiescent cosmology' proposed by Barrow \cite{Barrow}, such an isotropic and quiescent initial condition could be possible only if the equation of state for high density matter at the early universe represented the stiff fluid with $p = \rho$, so that an initially isotropic and homogeneous universe would be a probable one with the desired stability.

Extending his studies on the inhomogeneous models, Bonnor \cite{Bonnor00} investigated the arrow of time in the Szekeres cosmological models using the method similar to the one used in \cite{Bonnor0}. Here again he used an invariant depending on the Weyl and Ricci tensors, and introduced the definition of a vector representing the entropy flux of the gravitational field. Soon after that, Tod \cite{Tod1} proposed the `FRW conjecture', which was intimately connected with the hypothesis of zero Weyl tensor at initial singularities. He showed that a perfect fluid cosmology with a barotropic equation of state (EoS) which has an isotropic singularity at which the Weyl tensor vanishes, is necessarily a FRW solution. The hypothesis of zero Weyl tensor at the initial singularity would then explain the isotropy of the universe as well as its homogeneity.

In a bid to examine whether the quiescent cosmology \cite{Barrow} was a viable one, Goode et al. \cite{GCW} considered the class II Szekeres model, which is a simple example of a spatially inhomogeneous model that admits an isotropic singularity, and showed that the dimensionless scalar defined in Eq. \eqref{11} vanishes as one approaches the isotropic singularity, thereby preserving the WCH. They also considered other dimensionless scalars constructed from the Riemann tensor and its covariant derivatives. They insisted that the model of quiescent cosmology along with Penrose's proposal of gravitational entropy may provide a viable alternative to inflationary models \cite{Guth} when formulated within the framework of isotropic singularities. That the expression \eqref{11} serves as a better estimator of GE was further confirmed in the paper \cite{RA1}, where it was verified that it exhibits the correct behaviour for the density perturbations in dust-filled models in the longitudinal gauge for a flat expanding universe (with perturbations only along the $z$ direction), as well as in spatially flat but expanding gravitational wave spacetime.

Grøn and Hervik \cite{GH1} investigated the evolution of different measures of GE in Bianchi type I and Lema\^{\i}tre–Tolman–Bondi (LTB) universe models, and found that the WCH remains secure in the non-local version of the conjecture. They \cite{GH1,GH2} showed that according to the classical field equations of Einstein, the quantity $P^2$ diverges at the initial singularity both in the case of the homogeneous, but anisotropic Bianchi type I cosmological models with dust and a cosmological constant, and also in the isotropic, but inhomogeneous LTB models. This meant that there were large anisotropies and inhomogeneities near the initial singularities in these models. They introduced a new quantity preserving the SLT, and investigated whether a quantum mechanical calculation of initial conditions for the universe based upon the Wheeler-DeWitt equation preserves the WCH or not. They considered two different representations for this purpose. The first one was a local version which failed to support the WCH. Then they constructed a non-local entity proportional to $P$ which would remain finite at the initial singularity even if $P$ diverges, thereby exhibiting favourable behaviour regarding the WCH, and reflected the tendency of the gravitational field to generate inhomogeneities. 	

Barrow and Hervik \cite{BH} studied the evolution of the Weyl curvature invariant in spatially homogeneous anisotropic universe models, and showed that the Weyl curvature invariant typically dominates the Ricci invariant at late times.
However, the definition of GE as the ratio of the Weyl square and the Ricci square faced problems in the case of a collapsing sphere of perfect fluid which conducts heat and radiates energy to infinity \cite{Bonnor}.

\subsection{Bel-Robinson tensor as a measure of the energy of gravitational fields}
Exploring the analogy between the electromagnetic and the gravitational fields, Bel \cite{Bel1,Bel2} found a four-index tensor, constructed with the Riemann tensor of the gravitational field, which is analogous to the energy-momentum tensor of the electromagnetic field. This tensor, called the Bel tensor, vanishes if and only if the Riemann tensor is zero and is covariantly conserved in vacuum. In the vacuum case, the Bel tensor is often called the Bel-Robinson tensor. This tensor was independently studied by Robinson \cite{3}, who found the property of its complete symmetry. Thus the Bel–Robinson tensor $T_{abcd}$ was discovered in ordinary Einstein gravity (GR) at $D=4$ dimensions, in the search for a gravitational counterpart of the usual matter stress-energy tensor $T_{\mu\nu}$.

In 1997 Bonilla and Senovilla \cite{BonSen} presented some new properties of the Bel and Bel-Robinson tensors which indicated their potential to be related to the gravitational energy-momentum tensor. For any arbitrary spacetime they found a decomposition of the Bel tensor in terms of the Bel-Robinson tensor and two other tensors, which they called the ``pure matter'' super-energy tensor and the ``matter-gravity coupling'' super-energy tensor. They proved that the pure matter super-energy tensor of any Einstein-Maxwell field is actually the ``square'' of the usual energy-momentum tensor. This, together with the fact that the Bel-Robinson tensor has dimensions of the square of the energy density, led them to the definition of the \textbf{\emph{square root of the Bel-Robinson tensor}}, which is a covariant symmetric traceless tensor with dimensions of energy density, such that its ``square'' gives the Bel-Robinson tensor. They proved that the well-defined traceless square root $t_{ab}$ of the Bel-Robinson tensor exists if and only if the spacetime is of Petrov type $O$, $N$ or $D$. Of these, the type $O$ spacetime is conformally flat, for which the Weyl tensor vanishes. For other Petrov types they showed that the Bel-Robinson tensor can be factored into terms that are either Coulomb-like (as in type $D$ spacetimes), wave-like (as in type $N$ spacetimes), or more complicated. While the tracefree square-roots are unique, the factorisation of the Bel-Robinson tensor into two different symmetric tracefree tensors is generally not unique. This implied that spacetimes for which the free gravitational fields are a mixture of wave-like and Coulomb-like fields may be more complicated, and the definition of an effective energy density for free gravitational fields in such cases could also be complicated. Subsequently, they illustrated a new property of the Bel-Robinson tensor which enabled them to give a very
simple proof of the causal propagation of gravity in vacuum and provided an invariant characterization for Petrov type N space-times \cite{BonSen1}.

In the year 2000, Senovilla presented a purely algebraic construction of super-energy ($s-e$) tensors for arbitrary fields in any dimensions \cite{Senovilla}. This construction was quite useful because the timelike component of these $s-e$ tensors are positive definite and satisfy the dominant property, which is a necessary mathematical requirement of an energy density. These tensors are helpful in the consideration of the causal propagation of the fields.

Senovilla included the classical Bel and Bel–Robinson tensors for the gravitational fields in the general definition of super-energy tensors. He showed that for any field, there is a hierarchy of `$(super)^k$-energy' tensors and its derivatives that exist up to a given order related to $k$. This means that the $s-e$ tensors can be organized in a hierarchy of $(super)^k$-energy levels. Consequently, the Bel–Robinson tensor can be compared with the corresponding super-energy tensors of other physical fields. The correct physical units for any given $(super)^k$ energy density turns out to be those of energy density times $ 1/L^{2k}$. Moreover, the `strength' of a physical field at points where its energy density vanishes but such that every neighbourhood of it contains the physical field, requires the concept of the $(super)^k$-energy. It is for this reason that the Bel–Robinson tensor arises \emph{naturally} in General Relativity, where the energy density of the gravitational field can always be made to vanish at any point by appropriate choice of the coordinate system on account of the equivalence principle. Finally, the most important point is that all $s-e$ tensors give rise to divergence-free currents if the field generating them is isolated (no other fields interact with it), and these currents can be \emph{combined} to produce \emph{divergence-free currents} mixing the different fields.

Acquaviva and other authors \cite{AKS} analysed the credibility of the square root of the Bel-Robinson tensor as a proposal for the gravitational energy–momentum tensor. As this ``square root'' is constructed completely from the Weyl component of the Riemann tensor, it encodes the geometric properties of the free gravitational field. The authors utilized the general decomposition of any energy–momentum tensor to examine the thermodynamic interpretation of these geometric quantities. They found that although the matter energy-momentum remains conserved due to Einstein field equations, but the square root of the Bel-Robinson tensor is not necessarily conserved. The deviation from conservation is not an
observer-dependent effect. The authors asserted that such intrinsic dissipation could be related to a transfer of energy between
the macroscopic gravitational field and its underlying microscopic degrees of freedom.

\subsection{Gravitational entropy proposals based on the Bel-Robinson tensor}
There were several attempts to define the gravitational entropy (GE) on the basis of the Bel-Robinson tensor, and also in terms of the Riemann tensor and its covariant derivatives. Pelavas and Lake \cite{PL} showed that Eq. \eqref{11} does not serve as a suitable candidate for gravitational entropy along the homothetic trajectories of any self-similar spacetime.  They introduced a class of ``gravitational epoch'' functions which were dimensionless scalars, one of which was built from the Riemann tensor and its covariant derivatives only, and was denoted by $P$. They also suggested other alternative functions involving the Bel-Robinson tensor and analysed whether such functions could be regarded as gravitational entropy function or not.

Pelavas and Coley \cite{PC} discussed whether a suitable dimensionless scalar function might serve an acceptable candidate for the GE. They explicitly considered the Szekeres and Bianchi type VI$_h$ models that admit an isotropic singularity, and discussed other possible GE functions, including an appropriate measure of the velocity dependent Bel-Robinson tensor.

Using the fluid 4-velocity $u^a$, they constructed the positive scalar $W$ from the Bel-Robinsor tensor $T_{abcd}$ through the following prescription:
\begin{equation}\label{12}
 W= T_{abcd} u^a u^b u^c u^d,
\end{equation}
which has the same units as $C_{abcd} C^{abcd}$. Then they obtained a dimensionless scalar by normalizing with the help of the square of the energy density $\mu = T_{ab} u^a u^b$, i.e.,
\begin{equation}\label{13}
\tilde{P} = W/\mu^2 .
\end{equation}
They also defined a dimensionless ratio of differential invariants, for example
\begin{equation}\label{14}
P_1=\nabla_a C_{bcde} \nabla^a C^{bcde}/\nabla_a R_{bc} \nabla^a R^{bc} .
\end{equation}

\bigskip

There are also few other variants of the gravitational entropy function derived from the Bel-Robinson tensor, one of which will be discussed in details in the subsequent text.

\section{SUBSEQUENT FORMULATIONS OF GRAVITATIONAL ENTROPY AND RECENT DEVELOPMENTS}

An interesting approach to handle the problem of gravitational entropy from a phenomenological point of view was proposed in 2008 by Rudjord et al. \cite{RGS} and subsequently expanded by Romero et al. in 2012 \cite{RTP}. Their objective was to test whether the GE computed according to their proposal in the case of certain black holes and other cosmological models, agrees with the WCH or not. The Rudjord proposal is directly inspired from the fact that black hole entropy is related to its geometry. Rudjord et al. showed that the Weyl scalar alone is not a good measure of GE. The same thing is true for the ratio of the Weyl scalar squared to the squared Ricci tensor. They explicitly showed that their measure of GE in terms of the ratio of the Weyl scalar to the Kretschmann scalar serves as a good measure and reproduces the Hawking-Bekenstein entropy for black holes. Romero et al. applied this proposal to several other systems of black holes and worm holes, validating and extending the Rudjord et al. proposal of GE. This proposal due to Rudjord et al. and Romero et al. will be referred to as the ``Weyl proposal'' (which yields a geometric measure) of GE in the remaining part of this article.

An important proposal of GE was offered by Clifton et al. \cite{CET}, who introduced a robust measure of GE based on the square root of the Bel-Robinson tensor, since this ``square root'' was already identified as a measure of the gravitational energy–momentum tensor \cite{BonSen}. Their idea was motivated by the considerations of relativistic thermodynamics, and the measure proposed by them had the interpretation of the effective super-energy-momentum tensor of free gravitational fields. According to the requirements listed in Section 2 of their paper, for a viable proposal of GE, it is not necessary for the GE to increase with the Weyl curvature, although that was what Penrose had envisaged. That's because such a restriction is too strong to be satisfied \cite{ES}, as a result of which several other measures of GE were also proposed following Penrose's hypothesis, which were only suitable functions of the Weyl curvature. Thus an increasing Weyl curvature is not a necessary, but only a sufficient condition for an increasing GE function. The spacetime should have a viable non negative measure of GE which increases monotonically with time. The specific form of the measure of CET entropy depended on the nature of the gravitational field --- Coulomb-like or wave-like. However, this definition of GE is only valid for General Relativity, where the Bel-Robinson tensor can be defined in this way. The authors in \cite{CET} explicitly showed that their proposal reproduces the Hawking-Bekenstein entropy for stationary BHs. They considered scalar perturbations of a FLRW geometry and found that the GE behaved like the Hubble weighted anisotropy of the gravitational field, so that it increases as structure formation takes place. However, the FLRW metric is conformally flat \cite{Ellis}, and so the Bel–Robinson tensor vanishes. Naturally the gravitational epoch function $W$ derived from the Bel–Robinson tensor, also vanishes, and so does the corresponding measure of GE. It may be noted that a viable measure of GE is identified with the presence of inhomogeneity, which requires the presence of both anisotropy and a non-zero $W$. For the non-perturbative inhomogeneous LTB models it was found that the GE increases as long as the expansion weighted value of gravitational energy density $\rho_{grav}$ increases. In the subsequent text, we will refer to this formulation as the Clifton-Ellis-Tavakol proposal or the ``CET proposal'' (which is purely a thermodynamic measure) of GE.

Among the two measures of gravitational entropy mentioned above, the \textbf{Weyl proposal} is by far the most suitable geometric measure, whereas the \textbf{CET proposal} is a robust proposal based on relativistic thermodynamics. The application of the CET proposal provides unique gravitational entropies for spacetimes of Petrov type $D$ and $N$ only, as they are derived from the square root of the Bel–Robinson tensor, whereas the geometric method can be used in almost all kind of spacetimes.

To examine the status of the WCH in inhomogeneous Universe, Li et al. \cite{LBHMS} studied two scalar estimators to quantify the deviations from a homogeneous and isotropic spacetime. These were the relative information entropy and a Weyl tensor invariant, and indicated their relation to the general averaging problem in the inhomogeneous Universe. They found that these two quantities could be linked via the kinematical backreaction of a spatially averaged universe model, when calculated up to second order in standard cosmological perturbation theory. With regard to the WCH, they found that not only was the Weyl curvature growing monotonically, but a specific combination of it with the kinematical backreaction was also growing monotonically, and this contributed to the observed acceleration of the Universe. Very soon Sussman \cite{Sussman} introduced the formalism of weighed scalar average (`q-average' formalism) for studying the theoretical properties and the dynamics of spherically symmetric LTB dust models, and explored the applicability of this formalism to a GE function proposed by Hosoya and others \cite{HBM} (the HB proposal). It was demonstrated that a positive entropy production follows from a negative correlation between fluctuations of the density and Hubble scalar, fulfilling the HB proposal in various LTB models and regions.

In a paper, Sussman and Larena \cite{SL1} examined the CET proposal and the HB proposal, along with a variant of the HB proposal, in the case of the generic LTB dust models. The conditions for the growth of GE for all three proposals were found to be directly related to a negative correlation of similar fluctuations of the energy density and Hubble scalar. This correlation was evaluated locally for the CET proposal, but for the two HB proposals it had to be evaluated in a non-local domain dependent manner. The results for all three proposals was clubbed together as follows: (i) the GE grows whenever the density growing mode is dominant; (ii) all ever-expanding hyperbolic models reach a stable terminal equilibrium characterized by an inhomogeneous maximum for the GE in their late time evolution; (iii) regions with decaying modes and collapsing elliptic models exhibit unstable equilibria associated with a minimum value of the GE; (iv) near singularities the GE function in the CET proposal diverges while those in the two HB proposals converge; finally, (v) the CET entropy function converges for all models in the radial asymptotic range, whereas the HB entropy functions only converge for models asymptotic to a FLRW background. Their conclusion was that the notion of GE is theoretically robust, and can also be applied to other generic spacetimes. In another work \cite{SL2}, these authors studied the evolution of the CET gravitational entropy for the local expanding cosmic CDM voids $\sim 50-100$ Mpc size, whose dynamics is determined by the LTB dust models, using a non-perturbative approach. Individually, Sussman \cite{Sussman1} studied the CDM void configurations compatible with basic observational constraints, to show that the CET gravitational entropy grows from post-inflationary conditions towards a final asymptotic value in a late time fully nonlinear regime described by the LTB dust models.

P\'{e}rez and Romero \cite{PerezRomero} studied the two estimators of GE based on scalars constructed from the Weyl tensor, in Kerr spacetime. Based on a simple choice of a timelike unit vector $u_a$ and a spacelike unit vector $z^a$ that determines the Weyl principal tetrad, they calculated the gravitational energy density, temperature, and the GE of a Kerr BH according to the CET estimator. They proved that with the simplest coordinate choice, the gravitational entropy failed to reproduce the Bekenstein-Hawking entropy of a Kerr black hole. Their result seemed to be a general one due to the fact that the innermost region of the Kerr spacetime is not folliable and time-orientable. However, the Weyl estimator did produce the desired results for Kerr and Kerr-Newman spacetimes.

Bolejko and Stoeger \cite{BS} studied the problem of GE along with the intermediate homogenization of inhomogeneous cosmological models. They considered five possible measures of GE beginning with (i) the standard canonical definition, i.e., the ratio of the Weyl to the Ricci curvature, (ii) the integrated version of the standard canonical definition, (iii) the canonical definition multiplied by the square root of the determinant of the spatial metric following the recommendation of Gr{\o}n and Hervik \cite{GH1}, (iv) the definition suggested by Hosoya et al. \cite{HBM} and subsequently used by Sussman \cite{Sussman}, and finally, (v) followed the definition proposed by Clifton et al. \cite{CET}. All definitions of entropy examined by them yielded decreasing gravitational entropy during the \textbf{\emph{homogenization process}}.

Marozzi et al \cite{MUUC} calculated the GE function of the large scale structures of the universe in the linear regime, where it can be described by the perturbed FLRW spacetime. In this case, the GE arose from the averaging made over an extended region and explains the formation of large-scale structure in the Universe. Their results agreed well with those obtained in \cite{SL2} for the linear regime of the LTB evolution, thereby indicating that there is a connection between the local physics and the large scale linear regime. The idea that emerged from their investigation is that we must adopt a suitable notion of GE in order to account for the growth of structures in the Universe. From statistical mechanics we know that the entropy of a system is related to the total number of microstates corresponding to a given macrostate. When translated into the language of cosmology, this implies that the notion of GE should be related to the number of small scale inhomogeneities that are consistent with the dynamics of a given homogeneous universe.

By choosing a minimal void LTB model for their analysis, Mishra and Singh \cite{MS} examined whether inhomogeneous cosmological models could be motivated on a thermodynamic basis. They studied several definitions of GE and found that the definition of the GE function proposed by Wainwright and Anderson \cite{WA} exhibits satisfactory thermodynamic behavior in the case of inhomogenous cosmologies.

Bolejko \cite{Bolejko} demonstrated that the notion of GE of the Universe is consistent with the cosmic no-hair conjecture (that a universe dominated by dark energy should asymptotically approach a homogeneous and isotropic de Sitter state). He used a simulation that approximated the universe by a model that represents irrotational, non-viscous, and insulated dust, with vanishing magnetic part of the Weyl curvature. The simulation was run to evolve the universe from $t=$ 25 Myr to $t=$ 1000 Gyr after the big bang. It was found that a universe with a positive cosmological constant and non-positive spatial curvature does approach the de Sitter state, but it still keeps generating the gravitational entropy.

Acquaviva et al \cite{AEGH} examined the thermodynamic properties of BH formation through gravitational collapse in the framework of 1 + 1 + 2 semitetrad covariant formalism, and proved that in the case of the Oppenheimer–Snyder–Datt collapse, the change in gravitational entropy outside a collapsing object is related to the variation of the surface area of the collapsing object, even before the formation of any horizon. Thus the Bekenstein–Hawking entropy of the BH end state could be correlated with the variation of the GE in the empty region outside the collapsing object. The authors showed that the CET measure of gravitational entropy yields the Bekenstein–Hawking entropy of the BH end state during a spherically symmetric collapse of a massive star with Schwarzschild spacetime as the exterior, and this entropy is the difference between one-fourth of the initial area of the collapsing star and the net increase in the entropy of the exterior region in infinite collapsing time.

Lima et. al. \cite{LNP} calculated the GE function for wormholes with exotic matter and in galactic halos in the framework of the Weyl proposal of gravitational entropy. They found that the gravitational entropy and entropy density of these wormholes in
regions near their throats are indistinguishable for objects of same throat, even if they are described by different metrics and by distinct energy-momentum tensors. Further, the GE function for both wormholes were found to exhibit a behaviour similar to the Hawking-Bekenstein's entropy on the surface of nonrotating and null charge black holes.

Gregoris and co-authors \cite{gregoris} considered a class of inhomogeneous spherically symmetric spacetimes supported by chameleon massless scalar field that exhibit anisotropic shearing effects, with its evolution driven by a stiffened fluid thereby modelling the early Universe. They showed that, throughout the entire evolution, the matter curvature dominates over the Weyl curvature, but the CET gravitational entropy remains non-negative and monotonically increasing. However, the definition of GE adopted by them depended not only on the Weyl curvature but also on the shear tensor that led to the increase in the magnitude of the GE function in their model. The shear played an important role because not only that it affects the dynamics of the electric part of the Weyl tensor resulting in a change in gravitational energy density, but also influences the gravitational temperature.

Guha and Chakraborty \cite{GuhaChak} investigated whether the estimator of GE in the Weyl proposal could be applied to the case of accelerating BHs. They found that such a proposal produced satisfactory results for the accelerating BHs and charged accelerating BHs, barring the case of the rotating charged accelerating metric. Chakraborty et al. \cite{CGG1} examined the validity of the CET definition of gravitational entropy for some well-known isotropic and anisotropic cosmologies, namely, the FLRW model, LRS Bianchi I model, Liang model, class II Szekeres model, and the Bianchi VI$_h$ model. They calculated the relevant functions like the normalized epoch function, gravitational energy density, gravitational temperature and the gravitational entropy for these models, and checked that in the vicinity of the initial singularity, the ratio of energy density of free gravity to that of matter density goes to zero, validating the WCH.

Subsequently, the same authors examined the validity of the CET and Weyl proposals of gravitational entropy in the context of traversable wormhole solutions of the Einstein field equations \cite{CGG2}. Since the proposals for a suitable measure of GE are still evolving, therefore, the investigations of the GE function for such models was necessary to yield a thermodynamic perspective of the physical reality of these solutions. The thermodynamic stability of these solutions depends on the well behaved nature of the GE function. Otherwise, traversability cannot be guaranteed. The authors examined the GE function of the concerned wormhole solutions both in terms of the Weyl proposal, as well as the CET proposal. Considering some of the Lorentzian traversable wormholes along with the Brill solution for NUT wormholes and the AdS wormholes, the authors evaluated the gravitational entropy and the gravitational entropy density for these systems. Both the Weyl and the CET proposals yielded consistent measures of GE for several of the wormhole solutions that were considered.

Very recently Piza\~{n}a and co-authors \cite{PSH} investigated the CET proposal for the Szekeres class I models, which are much less idealized spactimes that can describe the simultaneous evolution of several types of structures like overdensities and voids, all of which are located on specific spatial positions in an asymptotic $\Lambda$CDM background. They found that a negative sign of the product of the density and Hubble expansion fluctuations constitute the necessary and sufficient conditions for a positive CET entropy production. To determine the viability of this theoretical prediction, they did a numerical estimate of the CET entropy production for two elongated over dense regions surrounding a central spheroidal void, all evolving jointly from initial linear perturbations at the last scattering era into the present Mpc-size CDM structures. They proved that the CET entropy production is positive for all times after the last scattering at the spatial locations where structure formation takes place and the density growing mode dominates.

Using only the Hamiltonian constraint \cite{Roupas1}, it was shown in the framework of GR that the gravitational potential inside any kind of matter in a static spacetime, can be derived from the maximum entropy principle, without invoking the Einstein equations or any other principle. This indicated an implicit thermodynamic sector within GR, and designated a thermodynamic property that any other alternative theory of gravity may also need to satisfy. The same author \cite{Roupas2} demonstrated that regular BHs in GR containing a de Sitter core correspond to a degenerate infinite spectrum of solutions, assuming quantum indeterminacy of the localization of the horizon, which behaves as an anisotropic fluid shell. All states of the cosmological black hole spectrum have the same energy and entropy, resembling quantum degeneracy. This entropy is the fluid entropy, and it recovers the Bekenstein–Hawking entropy if the Tolman temperature of the fluid is identified with the temperature of the cosmological horizon, fusing the cosmological and BH horizons in a single dual horizon.

\section{TWO USEFUL PROPOSALS OF GRAVITATIONAL ENTROPY}

By now we have understood that among the various measures of gravitational entropy proposed till today, some are purely the square of the Weyl scalar, whereas some are ratios involving it, and some others are derived from it through a suitable prescription.
In this section we will discuss in details the Weyl proposal and the CET proposal of gravitational entropy, both of which have been applied to physical situations by several authors. As chronologically the Weyl proposal came first and then the CET proposal, we will follow the same sequence in the following exposition.

\subsection{The Weyl proposal of gravitational entropy}
Let us begin with a brief description of the Weyl proposal given in \cite{RGS} for the determination of gravitational entropy.

The entropy of a BH is described by the surface integral
\begin{equation}\label{15}
S_{\sigma}=k_{s}\int_{\sigma}\mathbf{\Psi}.\mathbf{d\sigma} ,
\end{equation}
where $\sigma$ denotes the surface of the horizon of the BH, and for static spherically symmetric spacetimes, the 3D vector field $\mathbf{\Psi}$ is assumed to possess only a radial component given by
\begin{equation}\label{16}
\mathbf{\Psi}=P \mathbf{e_{r}},
\end{equation}
with $ \mathbf{e_{r}} $ as a unit radial vector. Here $P$ represents either $P_1$ or $P_2$. The scalar $P_1$ is defined in terms of the Weyl scalar ($W$) and the Krestchmann scalar ($K$) in the form
\begin{equation}\label{17}
P_1^2=W/K=C_{abcd} C^{abcd}/R_{abcd} R^{abcd} .
\end{equation}
This is a purely geometric measure of GE, but it contains the ratio of the Weyl to the Krestchmann scalar, and hence it nicely encompasses the curvature dynamics as a whole.

In this method, in order to find the gravitational entropy, one has to do computations in a 3-space. Hence, the spatial metric  $h_{ab}$ is defined as
\begin{equation}\label{18}
h_{ij}=g_{ij} - \frac{g_{i0} g_{j0}}{g_{00}} ,
\end{equation}
where $g_{\mu\nu}$ is the corresponding 4-dimensional space-time metric and Latin indices denote spatial components, $i,j = 1,2,3$.

So the infinitesimal surface element is given by
\begin{equation}\label{19}
d\sigma=\frac{\sqrt{h}}{\sqrt{h_rr}} d\theta d\phi.
\end{equation}

In the case of wormholes which do not have any horizons, one has to switch into the gravitational entropy density, $s$. In order to evaluate the entropy density, one has to imagine an enclosed hypersurface and use the Gauss's divergence theorem to obtain that
\begin{equation}\label{20}
s = k_s |\nabla \cdot \mathbf{\Psi}| .
\end{equation}
Since $\mathbf{\Psi}$ has only radial component, one obtains
\begin{equation}\label{21}
s=\frac{k_s}{\sqrt{-g}} |\frac{\partial}{\partial r} (\sqrt{-g} P_i )|.
\end{equation}
If there is an angular component in the vector field $\mathbf{\Psi}$, as in axisymmetric spacetimes considered by \cite{RTP}, then one has to use the  modified definition of $\mathbf{\Psi}$, to calculate the gravitational entropy density for axisymmetric space-times, which is then given by 
\begin{equation}\label{22}
s=\frac{k_s}{\sqrt{-g}} |\frac{\partial}{\partial r} (\sqrt{-g} P_i ) + \frac{\partial}{\partial \theta} (\sqrt{-g} P_i ) |.
\end{equation}
To evaluate this entropy density, one has to use the following measure proposed in \cite{RTP} for the expression of $P_i$ in the case of metrics having nonzero $g_{t\phi}$ component:
\begin{equation}\label{23}
P_2=C_{abcd} C^{abcd}.
\end{equation}

\subsection{The CET proposal of gravitational entropy}

The CET proposal \cite{CET} deals with the entropy of the free gravitational field. In this formulation, the gravitational entropy is defined from an ``effective'' or ``super energy-momentum'' tensor $\tau_{ab}$ for the free gravitational field (a suitable geometric field associated with the Weyl tensor). Being constructed exclusively from the Weyl part of the Riemann tensor, this $\tau_{ab}$ encapsulates the geometric properties of free gravitational fields in terms of scalars describing null geodesic congruences. To establish the validity of this proposal, the authors have shown that it reproduces the Hawking-Bekenstein entropy when applied to BHs, and its entropy production rate, $S_{grav}$, is always non-negative. They have listed the requirements for a satisfactory definition of gravitational entropy as follows: (i) it should be non-negative, (ii) it should vanish if and only if the Weyl tensor vanishes, (iii) it should provide a measure of the local anisotropy of the gravitational field, and hence (iv) should reproduce the known results for black hole entropy, and finally, (v) it should increase monotonically with structure formation in the universe. The authors constructed the definition of gravitational entropy from the Bel–Robinson tensor, and from semi-classical notions of the temperature of a gravitational field.

The Bel–Robinson tensor is defined in terms of the Weyl tensor as \cite{Bel1,Bel2,robinson}
\begin{equation}\label{def1}
T_{a b c d} \equiv \frac{1}{4} \left( C_{e a b f}C^{e \; \; \; \; f}_{\; \; c d \;
\;}+C^{*}_{\; e a b f}C^{* \; e \; \; \; \; f}_{\; \; \; \; \; c d
\; \;} \right),
\end{equation}
where $C^{*}_{\; abcd}= \frac{1}{2} \eta_{abef} C^{ef}_{\phantom{ef}cd}$ is the dual of the Weyl tensor. This tensor is symmetric, tracefree, and covariantly conserved in vacuum (or in the presence of the $\Lambda-$term in cosmology). The factor of $1/4$ was included to give a natural interpretation of the Bel-Robinson tensor in terms of the Weyl spinor \cite{PR}. The gravito-electromagnetic properties of the Weyl tensor, and the 1+3 decomposition of the equations were used to express the gravitational ``super energy density'' function $W$ as follows:
\begin{equation}\label{28}
W= T_{abcd} u^a u^b u^c u^d = \frac{1}{4} (E_a^b E_b^a + H_a^b H_b^a ).
\end{equation}
The quantities $E_{ab}$  and $H_{ab}$ are the `electric' and `magnetic' parts of the Weyl tensor $C_{abcd}$ respectively (by analogy with the Maxwellian decomposition of the electric and magnetic fields as measured by $u^a$ observers in electromagnetism \cite{MB,Schouten}).

The authors constructed a second order symmetric traceless tensor $t_{ab}$ which is obtained from the algebraic ``square root'' of the fourth order Bel-Robinson tensor $T_{abcd}$, which is the only totally symmetric tracefree tensor that can be constructed out of the conformal Weyl tensor $C_{abcd}$. For the four-index tensor, $T_{abcd}$, it is possible to define the symmetric two-index ``square-root'', $t_{ab}$, as a solution to the equation (see \cite{Schouten})
\begin{equation}\label{31}
T_{abcd} = t_{\left(ab \right.}  t_{\left. cd \right)} -  \frac{1}{2}  t_{e \left(a\right.} t_b^e g_{\left.cd\right)} -  \frac{1}{4}  t_e^e  t_{\left(ab\right.}  g_{\left. cd \right)} + \frac{1}{24} \left(t_{ef} t^{ef} + \frac{1}{2} (t_e^e )^2 \right) g_{\left( ab \right.}  g_{\left.cd\right)}.
\end{equation}
For a given $t_{ab}$, this equation gives the unique symmetric tracefree four-index tensor $T_{abcd}$, but it is not true that there is a $t_{ab}$ for any arbitrary symmetric and tracefree four-index tensor. As the fourth order $T_{abcd}$ depends on the square of the Riemann tensor, therefore, it has the dimensions of $L^{-4}$ (where $L$ means length) \cite{BonSen}, making it necessary to take its ``square root''.

This $t_{ab}$ helps one to derive the ``effective'' or ``super energy–momentum tensor'' $\tau_{ab}$ of the free gravitational field. It is possible to compute other variables like gravitational energy density $\rho_{grav}$, gravitational pressure $p_{grav}$, anisotropic stresses $\Pi_{grav}^{ab}$, and heat flux $q_{grav}^a$ by contracting with the matter 4-velocity $u^a$ and the projection tensor $h_{ab} =u_a u_b + g_{ab}$. Subsequently one can arrive at a clear notion of gravitational entropy by analogy with the standard laws of fluid thermodynamics applied on the quantities associated with $\tau_{ab}$. A suitable definition of gravitational entropy is expected to be observer dependent covariant quantity. Further, the two quantities of gravitational entropy and matter entropy should together maintain the total of entropy for all fields as an extrinsic quantity.

It may be recalled that from the work of Bonilla and Senovilla \cite{BonSen}, it was known that the tracefree square root $t_{ab}$ of the Bel-Robinson tensor exists if and only if the spacetime is of Petrov type $O$, $N$ or $D$, of which the Weyl tensor vanishes for the conformally flat type $O$ spacetime. The algebraic decomposition of the Bel-Robinson tensor into a second order effective energy momentum tensor can also be done for other Petrov type spacetimes, but it is only in the Petrov type $D$ and $N$ spacetimes that the second order effective energy momentum tensor is unique \cite{BonSen}.

Thus the CET proposal is applicable only to Einstein's gravity as it can provide unique expression for gravitational entropy only for the Petrov type $D$ and type $N$ spacetimes. The type $D$ and type $N$ cases contain interesting examples that include all stationary black hole solutions, which are useful for checking the established definitions of gravitational thermodynamics, as well as the case of scalar perturbations of Robertson-Walker geometries, which are important in cosmology.

In the CET paper, the authors considered two types of gravitational fields: the ``Coulomb-like'' (Petrov type $D$) and the ``wave-like'' (Petrov type $N$) fields for which $\tau_{ab}$ reduces to expressions involving the Newman–Penrose conformal invariants $\Psi_2$ and $\Psi_4$.

\subsubsection{CET proposal for Coulomb-like fields}

For the Coulomb-like Petrov type $D$ fields (these have two double principal null directions), the authors derived the following $\tau_{ab}$ and the associated fluxes:
\begin{equation}\label{24}
8\pi \tau^{ab} = \alpha|\Psi_2 |[x^a x^b + y^a y^b - 2(z^a z^b - u^a u^b )] = \rho_{grav} u^a u^b + p_{grav} h^{ab} + 2q_{grav}^{\left(a \right.}  u^{ \left. b \right)} + \Pi_{grav}^{ab},
\end{equation}
\begin{equation}\label{25}
 8\pi \rho_{grav} = 2\alpha |\Psi_2 |,
\end{equation}
\begin{equation}\label{26}
 p_{grav}= q_{grav}=0,
\end{equation}
\begin{equation}\label{27}
8\pi \Pi_{grav}^{ab} = \frac{\alpha|\Psi_2 |}{2} (x^a x^b + y^a y^b - z^a z^b + u^a u^b ).
\end{equation}
Here, $\alpha$ is a positive constant which provides appropriate physical units, and [$u^a,x^a,y^a,z^a$] constitute an orthonormal tetrad defined as follows:
\begin{equation}\label{29}
m^a=\frac{1}{\sqrt{2}} (x^a - i y^a), \quad l^a = \frac{1}{\sqrt{2}}(u^a - z^a), \quad \textrm{and} \quad k^a= \frac{1}{\sqrt{2}}(u^a + z^a),
\end{equation}
where $x^a$, $y^a$ and $z^a$ are spacelike unit vectors, which constitute an orthonormal basis together with $u^a$. Using these, the entire metric can be rewritten in terms of the tetrads: $g_{ab}=2 m_{\left( a \right.} \bar{m} _{\left. b \right)} - 2 k_{\left(a \right.} l_{\left. b\right)}$, with $l^a$ and $k^a$ being aligned with the two principal null directions.

For these fields, the tracefree square-root is given by \cite{BonSen}
\begin{equation}\label{em}
t_{ab} = 3 \vert \Psi_2 \vert \left( m_{(a} \bar{m}_{b)} + l_{(a} k_{b)} \right),
\end{equation}
where $\Psi_2= C_{abcd} k^a m^b \bar{m}^c l^d$ is the only non-zero Weyl scalar.

Therefore in this scheme of the free gravitational field \cite{CET}, the effective gravitational energy density can be written as:
\begin{equation}\label{30}
8\pi \rho_{grav} = 2\alpha \sqrt{\frac{2W}{3}},  \qquad  |\Psi_2 |=\sqrt{\frac{2W}{3}}, \qquad  \textrm{with} \qquad  \rho_{grav} \geq 0.
\end{equation}
The relation between $\Psi_2$ and $W$ indicates that the important properties of the Bel–Robinson tensor are inherited
by the effective energy–momentum tensor $\tau_{ab}$, because $W$ is observer dependent and non-negative, and vanishes if and only if the Weyl tensor vanishes. As a result, the gravitational entropy, $S_{\rm grav}$, is non-negative, it vanishes if and only if $C_{abcd}=0$, and further it is evident that it serves as a measure of the local anisotropy of the free gravitational field if one takes into account the case of perfect fluids.

\subsubsection{CET proposal for wave-like fields}

For the wave-like (Petrov type $N$) fields, the authors considered plane-fronted transverse gravitational waves (detectable through various gravitational wave experiments). The geometry associated with these waves belongs to the Kundt’s class \cite{Kundt}, which are Petrov type $N$ solutions with vanishing Newman-Penrose scalar.

For Petrov type $N$ spacetimes, in which all four principal null directions are degenerate, and the gravitational field is wave-like, the trace-free square root can be written as \cite{BonSen}:
\begin{equation}\label{32}
t_{ab} = \epsilon |\Psi_4 |  k_a k_b,
\end{equation}
where $\Psi_4= C_{abcd} \bar{m}^a  l^b  \bar{m}^c  l^d$ is the only non-zero Weyl scalar in this case, and $k^a$ is aligned with the principal null directions. Taking the effective energy-momentum tensor of this type of gravitational field to be given by the solution to \eqref{31}, with a trace-free part given by \eqref{32}, leads one to the following $\tau_{ab}$ of the form
\begin{equation}\label{33}
8\pi \tau_{ab}  = \beta \left[ \epsilon |\Psi_4 |  k_a k_b  + f g_{ab}\right],
\end{equation}
where $\beta$ is a constant (may or may not be equal to $\alpha$ in \eqref{24}. In order to have energy conservation in vacuum in this case, it is sufficient to take $f=\lambda_2=\textrm{constant}$. One can consistently and without loss of generality set the value of $\lambda_2 = 0$. The effective energy density, pressure, momentum density, and heat flux of this effective fluid are then

\begin{equation}\label{34}
8\pi \rho_{grav} = \beta \left(\frac{\epsilon}{2} |\Psi_4 |\right),
\end{equation}

\begin{equation}\label{35}
8\pi p_{grav}  = \beta \left(\frac{\epsilon}{6} |\Psi_4 |\right),
\end{equation}

\begin{equation}\label{36}
8\pi \Pi_{grav}^{ab} = - \beta \left[\frac{\epsilon}{6} |\Psi_4 |(x^a x^b + y^a y^b - 2z^a z^b )\right],
\end{equation}
and
\begin{equation}\label{37}
8\pi q_{grav}^a = \beta \left(\frac{\epsilon}{2} |\Psi_4 |\right) z^a.
\end{equation}
A set of orthonormal basis vectors have been used in the expressions for $\Pi_{grav}^{ab}$, and $q_{grav}^a$. It is possible to identify that $|\Psi_4 | = \sqrt{4W}$. The convenient choice is $\epsilon = +1$, so that, $\rho_{grav}\geq 0$.

The thermodynamic quantities for the wave-like gravitational fields then become

\begin{equation}\label{38}
8\pi \rho_{grav} = \beta \sqrt{4W}, \qquad  \textrm{and} \qquad  p_{grav}= \frac{1}{3} \rho_{grav},
\end{equation}
along with
\begin{equation}\label{39}
\Pi_{grav}^{ab} = - \beta \sqrt{4W} \left[ \frac{(x^a x^b+y^a y^b-2z^a z^b )}{48 \pi} \right], \qquad \textrm{and} \qquad 8\pi q_{grav}^a = \beta \sqrt{4W} z^a.
\end{equation}
The effective fluid in this case therefore takes the form of radiation-like matter fields, with an equation of state $\gamma = p/\rho = 1/3$.

\subsubsection{Gravitational temperature and the rate of production of entropy in the CET proposal}

To calculate the gravitational entropy $S_{ grav}$ according to the prescriptions of thermodynamics, one must find the ``temperature'', $T_{ grav}$ associated with the free gravitational fields, which is possible only if one knows about the microscopic properties of the gravitational field. For that purpose, the CET proposal invoked the results of black hole thermodynamics, and quantum field theory in curved spacetimes.
Assuming that a thermodynamic treatment of the free gravitational field is analogous to that of standard thermodynamics, the CET proposal required the definition of this ``temperature'' to be defined point-wise (i.e., not necessarily defined for horizons only). This requirement reproduced the expected results from semi-classical calculations in the case of Schwarzschild and de Sitter spacetimes.

The gravitational analogue of the fundamental law of thermodynamics was assumed to be given by
\begin{equation}\label{first}
T_{ grav} d S_{ grav} =  dU_{ grav} + p_{ grav} dv,
\end{equation}
where $T_{ grav}$, $S_{ grav}$, $U_{ grav}$ and $p_{ grav}$ denote the effective temperature, entropy, internal energy and isotropic pressure of the free gravitational field, respectively, and $v$ is the spatial volume.

Drawing analogy between the effective energy-momentum tensor $\tau_{ab}$ of the gravitational fields and the actual energy-momentum tensor of matter fields, the effective equation for the entropy production for perfect fluid matter in the case of Coulomb-like fields was obtained as
\begin{equation}\label{40}
T_{grav}  \dot{s}_{grav}  = (\rho_{grav}  v)^{\cdot} = - v \sigma_{ab} \left[\Pi_{grav}^{ab} + \frac{4\pi(\rho + p)}{3\alpha |\Psi_2 |}  E^{ab} \right].
\end{equation}

The independently defined local gravitational temperature at any point in space-time was given by the expression
\begin{equation}\label{41}
T_{grav}=\dfrac{|u_{a;b}l^{a}k^{b}|}{\pi}=\dfrac{|\dot{u_{a}}z^{a}+H+\sigma_{ab}z^{a}z^{b}|}{2\pi},
\end{equation}
where $z^a$ is a spacelike unit vector aligned with the Weyl principal tetrad, $\sigma_{ab}$ is the shear tensor, $\dot{u}_a = u^b \nabla_a u_b$ is the 4-acceleration,  $H = \Theta/3$ is the isotropic Hubble rate, and $\Theta \equiv \tilde{\nabla}_c u^c = h_c^b \nabla_b u^c$  is the isotropic expansion scalar. However, there may also be alternative definitions of temperature, as the above definition is \emph{\textbf{ad hoc}}.

In the case of wave-like fields, the equation describing the rate of entropy production in presence of perfect fluid matter reduced to the form
\begin{equation}\label{42}
T_{grav} \dot{s}_{grav} = (\rho_{grav} v)^{.} + p_{grav} \dot{v} = - v \left[g_{ab} q_{grav}^{a;b} + \dot{u}_a q_{grav}^a + \sigma_{ab} \Pi_{grav}^{ab} \right] - \frac{2\pi v(\rho + p) \sigma_{ab}  E^{ab}}{\beta \sqrt{4W}} .
\end{equation}
The first term in square brackets on the right-hand side of this equation represents relativistic dissipation, while the second term is due to the ``heat flow'' into the free gravitational fields from the matter fields. This term vanishes if $\rho=p=0$, i.e. $p=-\rho$, or when $\sigma_{ab}=0$, or, $E_{ab} = 0$.

For general Petrov type $N$ space-times an additional term appears on the right-hand side of this equation:
$$+ \frac{4\pi}{\sqrt{2} \beta} \sqrt{4W} \, v k_{a;b} (x^a x^b + y^a y^b).$$
This extra term vanishes for plane wave geometries in Kundt's spacetimes, but is non-zero in general \cite{RT}.

It is evident that the shear plays a very important role, because not only does it affect the dynamics of the electric part of the Weyl tensor resulting in a change in gravitational energy density, but it is also contained in the gravitational temperature.

\section{Applications of the Weyl and the CET proposals in specific cases}

\subsection{Gravitational Entropy of Wormholes with Exotic Matter and in Galactic Halos in the Weyl proposal}

In this work \cite{LNP}, the authors studied the nature of evolution of the gravitational entropy near the throat of traversable wormholes formed by exotic matter and wormholes in galactic halos. They verified that the gravitational entropy and entropy density of these wormholes in
regions near their throats are indistinguishable for objects of the same throat, although they are described by different metrics and distinct energy-momentum tensors and characterized by different boundary conditions. Both the Weyl and Kretschmann invariants were different for the two wormhole metrics but the expressions for the GE in the Weyl proposal were very similar.

Unlike the case of Romero et al. \cite{RTP} who found the gravitational entropy density of wormholes in classic exotic matter, where the entropy
density was shown to be null at the throat, the authors in \cite{LNP} found that the gravitational entropy density diverges near the throat for both types of wormholes, which may be due to a non-trivial topology at the throat. Such a situation may be attributed to the Noether current for gravity and the local entropy flux density for these wormholes. Physically, entropy is associated with information, so the authors interpreted that a maximum flux of information could be carried through the throat of these wormholes. They also found that both wormholes are endowed with an entropic behaviour similar to Hawking-Bekenstein’s entropy of non-rotating and null charge black hole.

\subsection{Gravitational entropy of accelerating black holes in the Weyl proposal}

In the analysis of accelerating BHs \cite{GuhaChak}, the authors considered four cases: (i) non-rotating uncharged BH, (ii) non-rotating charged BH, (iii) rotating uncharged BH, and (iv) rotating charged BH. Accelerating BHs are more realistic, for instance, galactic collisions are common in the universe, and it inevitably leads to BH mergers accompanied by the production of gravitational waves \cite{POK}. In such situations, the BHs at the centre of these galaxies will be accelerating towards each other, although even an isolated BH may be in accelerated motion, since no BH can be gravitationally separated from the neighbouring masses. Further, a static BH may be considered as the limiting case of an accelerating BH.

In the case of the non-rotating (uncharged or charged) BH, the GE was found to be proportional to the area of the event horizon of the BH, as in the case of the Bekenstein-Hawking entropy. Further, the zeroes of the gravitational entropy density function were located at the acceleration horizon and at the event horizon. The entropy density diverged near the $r=0$ singularity, and increased inside the horizon, encountering some zeroes in between. For the rotating uncharged BH, the GE function was given by a simple expression and was easily evaluated at the inner and outer horizons ($r_{\pm}$). However, for the rotating charged BH, the expression was quite complicated, although it could be computed. The expression for the gravitational entropy density for the uncharged rotating accelerating BH and the charged rotating accelerating BH turned out to be lengthy and very elaborate. In both these cases, the gravitational entropy density was found to diverge at the ring singularity and vanished at the conformal infinity.

\subsection{Gravitational entropy of traversable wormholes in the Weyl Proposal}

Among the six different traversable wormholes considered for the analysis of gravitational entropy in \cite{CGG2}, three were static spherically symmetric cases (namely the Ellis wormhole, the Darmour-Solodukhin wormhole, and the exponential metric wormhole). All of these three yielded a well-behaved finite measure of gravitational entropy density in the Weyl proposal. The authors also studied the traversable NUT wormhole, traversable AdS wormhole, and the Maldacena-Milekhin wormhole.

For the traversable NUT wormhole, the gravitational entropy density was computed using both the functions $P_1$ and $P_2$, taking into account both the radial and angular components in the definition of the entropy density $s$. It was shown explicitly that for these wormholes, the quantity $P_2$ gives a viable measure of gravitational entropy density.

For the NUT wormhole in the AdS spacetime, the Weyl proposal was applied for both the definitions of $P_1$ and $P_2$. In both these cases, the radial and the angular contributions together gave a complete picture, but as these metrics have a nonzero $g_{t\phi}$ component, the measure given by $P_2$ (with both the radial and angular contributions) turned out to be a far more complete and viable measure of gravitational entropy density.

The traversable wormhole system proposed recently by Maldacena et al. was analyzed to examine the behaviour of the gravitational entropy function of a traversable wormhole connecting two oppositely charged magnetic blackholes. In this wormhole ansatz, the Weyl proposal yielded zero gravitational entropy density, so that the system appeared to be nonphysical from the thermodynamic perspective of gravitational entropy. 

\subsection{Gravitational entropies in LTB dust models in the CET proposal}

In their paper \cite{SL1}, the authors conducted a comprehensive study of the CET proposal for the generic LTB dust models (considering FLRW-like metric parametrization), and also examined two other proposals. The dynamical and geometric properties of these models were described by means of an initial value parametrization of the metric, together with a covariant representation of q-scalars (a formalism of weighed scalar average introduced in \cite{Sussman}), their fluctuations and perturbations (using the method employed in \cite{Sussman0}) expressed in terms of exact generalizations of the density growing and decaying modes of linear perturbation theory.

The condition for the growth of GE was directly related to a negative correlation of similar fluctuations of the energy density and Hubble scalar, a result which was not obtained in the original CET paper. This correlation was evaluated locally and not statistically as the entropy growth involved the energy density and the Hubble scalar expressed in the q-scalar formalism, and it was found that the GE grows whenever the density growing mode is dominant. The condition of GE production was both necessary and sufficient. All ever-expanding hyperbolic models were found to reach a stable terminal equilibrium configuration characterized by an inhomogeneous maximum for the GE in their late time evolution, and regions with decaying modes and collapsing elliptic models exhibited unstable equilibria associated with a minimum value of GE. Further, they found that near the singularities the GE function was diverging, whereas the GE function converged for all models in the radial asymptotic range.

From the asymptotic and qualitative study of the CET entropy they showed that the GE function $s_{gr}$ was increasing near the big bang singularity and throughout the full time evolution only in models in which the decaying mode was totally suppressed. It was shown by earlier workers \cite{WA_other} that LTB models with a suppressed decaying mode emerge from an isotropic big bang singularity, and converge in earlier times to a spatially flat Einstein-de Sitter model. This feature therefore points toward a significant correlation between the early time behaviour of the CET entropy and basic geometric features of the initial singularity, and early time evolution of the models.

The authors also analysed several other aspects in their paper, like the necessity of suppressing the decaying mode, the relation between GE and cosmological `homogenization', the behaviour of the CET entropy in presence of a nonzero $\Lambda$ term, and the stability and extensivity of the LTB models in terms of the behaviour of the CET entropy.

\subsection{Gravitational entropy of cosmic voids in the CET proposal}

In another work \cite{SL2} Sussman and Larena studied the evolution of the CET gravitational entropy for the \textbf{\emph{local}} expanding cosmic CDM voids $\sim 50-100$ Mpc size described as spherical under-dense regions with negative spatial curvature, with the dynamics determined by the LTB dust models asymptotic to three different types of FLRW background. The authors used a non-perturbative approach to examine the effect of negative curvature and $\Lambda$ term on the evolution of the CET gravitational entropy, $s_{gr}$, inside the expanding sub-horizon of these voids. They used analytic expressions to examine the asymptotic time behaviour of the rate of change of $s_{gr}$, and its line integral, and did a numerical study to determine their full evolution from linear initial conditions specified at the last scattering time. Both analytic and numerical analysis revealed that the late time CET entropy growth is determined by the amplitude of initial fluctuations of spatial curvature at the last scattering time. The  growth of CET entropy was found to vanish in the late asymptotic time range for all voids, but at a faster rate in voids with $\Lambda$CDM and open FLRW backgrounds. The CET gravitational temperature reduced to zero asymptotically if $\Lambda = 0$ and became asymptotically proportional to $\Lambda$ for voids in a $\Lambda$CDM background.

Sussman \cite{Sussman1} applied the CET proposal to spherical expanding cosmic CDM voids compatible with basic observational constraints, within the framework of nonperturbative GR, and showed that the CET gravitational entropy grows from post-inflationary conditions towards a final asymptotic value in a late time fully nonlinear regime described by the LTB dust models. A similar behaviour was observed for $\Lambda$CDM background with a positive $\Lambda$-term, although the nonzero $\Lambda$-term led to a significant suppression of entropy growth with the terminal equilibrium value reached at a faster rate, as the rate of entropy growth reduced to zero.

\subsection{Entropy of black hole end state during gravitational collapse from CET gravitational entropy}

The authors in \cite{AEGH} examined whether the CET measure of gravitational entropy reproduced the Bekenstein–Hawking entropy of a BH which was formed at the end state of gravitational collapse. They considered the spherically symmetric Oppenheimer–Snyder–Datt collapse, which describes the gravitational collapse of a dustlike ball in a Schwarzschild exterior (a Petrov type $D$ spacetime), that made it simple to determine the GE as calculated by a static observer even before the formation of the event horizon. They proved that the Bekenstein–Hawking entropy of the BH end state formed after an infinite time for the static observer, can be linked to the net monotonic increase in GE in the empty region outside the collapsing object, as the collapse ensued. Their result established the relation between the time-varying gravitational field during the continuous process of matter collapse to the thermodynamic property of the BH end state, for which the concept of gravitational entropy is well explained \cite{Page3}.

The authors used the framework of 1 + 1 + 2 semitetrad covariant formalism for locally rotationally symmetric (LRS) class II spacetimes, and recasted the equations describing the gravitational entropy in this formalism, to prove that in the case of the Oppenheimer–Snyder–Datt collapse, the change in GE outside a collapsing object can be related to the variation of the surface area of the collapsing object, even before the event horizon is formed. Thus the CET measure of GE yielded the Bekenstein–Hawking entropy of the BH end state, which was given by the difference between one-fourth of the initial area of the collapsing star and the net increase in the entropy of the exterior region in infinite collapsing time, during the spherically symmetric collapse of a fireball in a Schwarzschild vacuum.

\subsection{CET gravitational entropy for shearing massless scalar field spacetimes}

In this paper \cite{gregoris}, the authors considered a class of inhomogeneous universe models sourced by chameleon massless scalar field that followed the canonical formalism and exhibited anisotropic shearing effects, with the purpose of determining whether the Weyl curvature necessarily increases in any physically realistic universe. The cosmic fluid in the model had the interpretation of a ``chameleon field'' as its EoS parameter depended on energy. The authors examined the effect of spatial inhomogeneities and a cosmological shear, and constrained the model by imposing several realistic criteria in the sense that the fluid must have a positive energy density that would represent a stiff fluid so as to model the early Universe, the thermodynamic second law must be obeyed in the matter sector, and the total matter entropy must be bounded by the area of the dynamical apparent horizon. The chameleon scalar field was considered to model scalar field inflation and the shearing effects would take care of any residual shear that prevailed after inflation. Such an assumption was justified on the basis that in the homogeneous but anisotropic Bianchi I models with regular matter content, a shear term dominates the primordial evolution, which becomes negligible at later times.

A negative deceleration parameter and a time-decreasing Weyl curvature automatically followed from these conditions, and the matter curvature dominated over the Weyl curvature throughout the entire evolution, and hence the matter entropy was sufficient to preserve the GSLT on the dynamical apparent horizon. Thus, in spite of all the imposed restrictions, the WCH did not hold in this class of models. However, the spacetime shear continued to increase with the expansion of the universe, and the CET entropy remained non-negative and monotonically increasing due to the increasing shear. This indicated the firm basis of the CET definition of gravitational entropy, which not only depends on the strength of the Weyl curvature, but also on the magnitude of spacetime shear. This behaviour persisted even if the chameleon property of the scalar field was switched off. Thus, in an accelerated expanding universe with growing shear, although the Weyl curvature decreases, but the CET entropy increases. This might point towards a non-trivial relation between gravitational entropy and spacetime shear. Further, the authors argued that their model was capable of explaining the formation of primordial structures like the large quasar groups and galaxies, without the necessity of invoking any quantum modification to general relativity.

\subsection{CET gravitational entropy of some cosmological models}

The models which were considered for the analysis of CET gravitational entropy in the paper \cite{CGG1}, are the following: FLRW model, LRS Bianchi I model, Liang model, Class II Szekeres model, and Bianchi VI$_h$ model.

For the homogeneous and isotropic FLRW universe, the authors found that the GE vanishes because the spacetime is conformally flat, thereby supporting the WCH because the free gravitational field in this case does not carry any gravitational energy density while maintaining a finite gravitational temperature. In the LRS Bianchi I case, the GE increases monotonically with time if its Weyl curvature increases with time, but if there are matter sources in the spacetime which cause the Weyl curvature to decrease in course of time, then the GE function in the LRS Bianchi I spacetime will decrease over time, thereby violating the WCH. Thus, in order to have a non-negative monotonically increasing GE, it was necessary for the LRS Bianchi I spacetime to have monotonically increasing Weyl curvature. In the Liang, Szekeres and Bianchi VI$_h$ models, the GE goes to zero as one approaches the initial singularity and increases monotonically with non-negative value in course of time. They found that in each of these models, representing different phases of evolution of the universe, the expansion anisotropy increases as times elapses after the initial isotropic singularity, and it appeared that a correlation exists between the expansion anisotropy and the gravitational entropy. They have also shown in a general formalism, that the shear tensor is related to the Weyl tensor. Therefore the CET formalism in these models gives us a well behaved entropy measure which increases as structure formation progresses, resulting in an increase in anisotropy in the corresponding universes. These features make these models physically more realistic for describing the actual evolution of the universe. In all these models, the gravitational entropy vanishes at the initial isotropic singularity. Moreover in each of these models, the gravitational energy density and the temperature are well behaved throughout the evolution of the conformal time associated with the metric.

\subsection{CET gravitational entropy of traversable wormholes}

The spherically symmetric static wormholes (WHs) that were considered in the paper \cite{CGG2} are the equilibrium cases where the condition $\dot{S}_{grav} = u^a \partial_a S_{grav} = 0$  holds strictly. The variation of the local piecewise gravitational entropy of static spherically symmetric spacetimes was defined by 
\begin{equation}\label{45}
S_{grav} (r) = \int \left( \frac{\rho_{grav} v}{T_{grav}} \right) dr,
\end{equation}
with the volume element as $v = 4 \pi \sqrt{h}$, where $h$ is the determinant of the projection tensor $h_{ab}$. 
The function $\rho_{grav} v /T_{grav}$ in \eqref{45} was treated like a `density'. For the Ellis wormhole, although the gravitational energy density was nonzero according to the CET proposal, but the gravitational temperature turned out to be singular, making it impossible to compute the GE from the CET prescription. As the definition of the gravitational temperature is completely \textbf{\emph{ad hoc}} in the CET proposal, so a new definition of gravitational temperature was introduced to obtain the GE, which then exhibited the desired behaviour. The rate of radial variation of the CET GE, as well as the local piecewise CET GE was finite at the throat, only going to vanishing limits for particular choice of parameters.

For the Darmour-Solodukhin wormhole, the CET proposal yielded a finite viable measure of GE, but the rate of radial variation of GE and the local piecewise CET GE diverged at the throat of the WH. For exponential metric wormhole, viable measures of GE density could be obtained for the CET proposal. The total GE was calculated as a function of the radial coordinate. For the traversable NUT wormhole, the CET proposal did not provide any unique expression for GE as the metric is not strictly of Petrov type $D$.

For the three static spherically symmetric wormholes which are in equilibrium (i.e., the Ellis WH, the DS WH, and the exponential metric WH), the authors checked whether these spacetimes satisfy the Tolman law which says that thermal equilibrium can exist within a temperature gradient if a gravitational field is present there. It was found that all three cases satisfy the Tolman law, implying that they are in thermal equilibrium. However, if one considered the original definition of gravitational temperature as given in the CET proposal, then none of the wormholes would satisfy the Tolman law. In the case of the traversable AdS wormhole, the CET proposal was not applied because that would require a far more careful and elaborate treatment. Finally, for the Maldacena wormhole ansatz, the CET proposal yielded zero GE density. For the sake of completeness the authors studied the extremal magnetic BHs of this system and showed that the relevant functions, i.e., the gravitational energy density, the gravitational temperature, the ratio of curvature scalars and the GE density are continuous at the junction of the black hole and the wormhole. The important point that emerged is that the CET GE of the magnetized extremal BH is proportional to the area of the horizon, and therefore agrees with the Hawking-Bekenstein entropy of a black hole.

\subsection{CET gravitational entropy in Szekeres Class I Models}

Piza\~{n}a and co-authors \cite{PSH} investigated the CET proposal for the quasi-spherical Szekeres class I models in FLRW-like coordinates, which describe the combined evolution of arrays of arbirary types of structures that include overdensities and voids, all of which are located on selected spatial positions in an asymptotic $\Lambda$CDM background. The spherically symmetric limit of these models is the class of LTB dust models.

The dynamics of these models is better described in terms of quasilocal scalars (named as `q-scalars') \cite{SB} related to local covariant scalars by means of weighted average functions defined by 3-dimensional integrals along the rest frames. The authors showed how the Einstein field equations reduce to a first-order system in terms of the q-scalars and non-spherical fluctuations. They examined the connection between the exact dynamics of Szekeres models with that of gauge-invariant cosmological perturbations by comparing the notion of density contrast with a similar exact quasilocal density fluctuation. Using suitable covariant variables and their fluctuations, they found that the necessary and sufficient conditions for a positive CET gravitational entropy production is that the product of the density and Hubble expansion fluctuations should have a negative sign.

To determine the viability of their theoretical predictions they conducted a general numerical estimate of the CET entropy production for two elongated over dense regions (in a $\Lambda$CDM background) surrounding a central spheroidal void, all evolving jointly from initial linear perturbations at the last scattering era into the present Mpc-size CDM structures. They proved that the CET entropy production is positive for all times after the last scattering at the spatial locations where structure formation takes place and the density growing mode dominates. The CET gravitational entropy when $\Lambda > 0$ evolved towards asymptotic values proportional to an asymptotic gravitational temperature.

\section{The evolving horizon of gravitational entropy}

To have a better insight into the interpretation of black hole entropy and understand the underlying  physics, Gregoris and Ong \cite{GregorisOng} constructed a new measure of gravitational entropy density function in terms of the Cartan curvature invariants and the Newman-Penrose scalars. They considered minimally coupled scalar field theories, where the entropy (written as the surface integral over angular variables on a section of the horizon), behaves as Noether charge \cite{Wald1}. Applying Gauss' theorem and introducing the inward and outward null normals to the horizon that satisfy the Newman-Penrose normalization, they constructed the entropy in terms of a volume integral of the divergence of the sum of these normals. They showed that their method is applicable to all static black hole solutions in four- and five-dimensional GR whether they are vacuum solutions or not. Since this method could be generalized to higher dimensions, the authors discussed the physical interpretation of black hole entropy, which becomes a bit mysterious when one considers the volume integral over $r$ (a temporal coordinate inside the horizon of a Schwarzschild black hole). However, the method becomes difficult to implement whenever the area law gets modified. Therefore, their work suggests that gravitational entropy in some modified theories of gravity is a manifestation of different physical effects since it is necessary to choose different combinations of curvature quantities. 

Turok and Boyle \cite{T&B} offered a new explanation for the observed large scale flatness, homogeneity and isotropy of our universe based on general relativity and Hawking's prescription of calculating GE. The novelty of their proposal rested on the special type of boundary conditions for big bang singularities, enforcing the tenets of CPT-symmetry. On account of these boundary conditions, they could calculate the GE function for cosmologies which included radiation, dark energy and spatial curvature of both type of signs. Their results justified the explanation offered by them.

From the realization that the various entropies existing in literature, namely the Barrow, Loop Quantum Gravity, Rényi, Tsallis, Sharma-Mittal, and Kaniadakis entropies (which were proposed as alternative to the Bekenstein-Hawking entropy), share some common properties, including the fact that they reduce to the Bekenstein-Hawking entropy in a certain limit, the authors in \cite{NOF1} investigated two new generalized entropies that contain all these previous proposals as special cases. The four properties shared by all these previous proposals, which are minimal requirements for any alternative entropy proposal, are:
\begin{enumerate}
  \item Generalized third law: The Bekenstein-Hawking entropy $S$ diverges when the Hawking temperature $T_H \rightarrow 0$ and vanishes as $T_H \rightarrow \infty$. Therefore the generalized entropies should vanish when the Bekenstein-Hawking entropy does.
  \item Monotonicity: The previous entropy proposals were monotonically increasing function of the Bekenstein-Hawking entropy $S$.
  \item Positivity: The entropies were positive, just as $S$, because the number of states $e^S$ is greater than unity.
  \item Bekenstein-Hawking limit: All the previous entropies reduce to the Bekenstein-Hawking prescription $S$ in an appropriate limit.
\end{enumerate}

The authors applied the generalized entropy to the Schwarzschild black hole and to spatially homogeneous and isotropic cosmology, where it was shown that it can potentially describe inflation and/or holographic dark energy. The first of the two new proposals in this paper features six parameters, while the second one is a simplified version containing only three parameters. The simplified proposal was shown to have the potential of generating an effective cosmological constant, which can drive early universe inflation as well as the present acceleration depending on its magnitude, and at the same time alleviate the current Hubble tension afflicting the standard CDM cosmological model.

Subsequently, in a second paper \cite{NOF2}, these authors advocated for the new entropy construction in \cite{NOF1} as it exhibits interesting phenomenology when applied to Schwarzschild black holes and holographic dark energy in cosmology. In this paper, the authors proposed ways of making the entire thermodynamics consistent with alternative entropies. They determined alternative entropies and corresponding energies in spherical, static, and asymptotically flat spacetimes and in modified gravity. They also determined the temperatures corresponding to these alternative entropies.

It is known that the thermodynamics of apparent horizon bridges with the usual FLRW equation only for a special case where the matter field is given by a perfect fluid with the equation of state (EoS) parameter equal to -1. To extend beyond such an EoS, the authors in \cite{NOP} considered the modification of the Bekenstein-Hawking entropy by developing an entropy function that leads to the usual FLRW equations, for a general EoS of matter, directly from the thermodynamics of the apparent horizon. The newly developed entropy involved a correction over the Bekenstein-Hawking entropy and differs from the known entropies like the Tsallis, Rényi, Barrow, Sharma-Mittal, Kaniadakis, and Loop Quantum Gravity entropies. The authors examined how the Friedmann equations of the apparent horizon cosmology are modified if one began with a general entropy that depends on the Bekenstein-Hawking entropy, and found interesting cosmological consequences during the early and late stages of the universe.

\bigskip

From the above account it is surely evident that the existence of a viable measure of gravitational entropy strictly depends on its definition.

\section{CONCLUDING REMARKS}

The theory of relativistic thermodynamics is sophisticated and attractive one, although a first look may not reveal the connection between gravity and thermodynamics. The connection emerges from the concepts of black hole thermodynamics. The most striking similarity between black hole physics and thermodynamics lies in the behaviour of the black hole horizon area and entropy, because both quantities tend to increase with time. The geometric feature of black hole temperature and entropy leads one to believe that gravity may also be interpreted in the same way. Sakharov in 1967 proposed that relativistic gravity emerges from quantum field theory in the same way that hydrodynamics or continuum elasticity theory emerges from molecular physics. The work of Bekenstein and Hawking provided the definition of black hole entropy. The analogy between black hole physics and the thermodynamics of the Universe led to the conjecture of Space-time Thermodynamics, according to which one can apply the laws of thermodynamics on the cosmological horizon, assuming the Universe to be a system bounded by a causal horizon.

According to the second law of thermodynamics, the total entropy of the Universe should increase with time. But the Universe was born in a homogeneous state and later on, small density fluctuations appeared due to the effect of gravity, leading to the formation of stars and galaxies. Observations of the cosmic microwave background radiation indicate that the early universe was in a state of thermal equilibrium, which should therefore correspond to a state of maximum entropy, instead of a state of low entropy. This is the initial entropy problem. The problem may be circumvented if one remembers that the early universe was filled with radiation streaming in all directions, and was devoid of matter, since structure formation had not kicked off because the gravitational degrees of freedom were not yet excited. Thus the maximum value of entropy in the initial universe was contributed by radiation only. The total entropy should also take into account the gravitational entropy, which was absent or negligible in the early universe. The early universe was characterized by local thermal equilibrium, but global thermal equilibrium did not exist as the universe was expanding. All these considerations necessitated a suitable proposal of gravitational entropy which should have a small value for the homogeneous distribution in the initial universe and should increase with time as more and more gravitational degrees of freedom got excited leading to structure formation. The development of inhomogeneities led to the increase of entropy, as gravitational energy got converted to thermal energy and electromagnetic radiation were also emitted. However, the process of formation of black holes leads to the maximum increase of gravitational entropy, with the maximum value given by the Bekenstein-Hawking entropy.

As the Weyl curvature is zero for a homogeneous distribution, Penrose conjectured that gravitational entropy should be related to the Weyl curvature of a spacetime. Several measures of gravitational entropy have been introduced after Penrose’s conjecture, and the role of tensors in all these measures \emph{\textbf{can hardly be undermined}}. In spite of these, there are differences between purely geometric formulations in terms of tensors and calculations based on physical principles. Thus, the total gravitational entropy given in terms of the volume integral of the divergence of the square root of the ratio between the Weyl and Kretschmann scalars is usually different from the Bekenstein-Hawking entropy of the horizons except for specific spacetimes. Only in the case of the Schwarzschild spacetime, the entire entropy of the horizon originates from the gravitational entropy arising out of the inhomogeneity of the gravitational field. The conformally flat spacetimes have zero Weyl curvature, and have vanishing gravitational entropy. In simple terms, we can say that a wide range of different situations arise depending on the nature of the spacetime. 

This article is an effort to weave together the development of the concept of generalized entropy as appearing in the generalized second law of thermodynamics, and the concept of gravitational entropy motivated by the Weyl curvature hypothesis and the principles of black hole thermodynamics. We took a tour of the various proposals of gravitational entropy, with particular emphasis on two specific proposals, namely the Weyl proposal and the CET proposal, and reviewed some of the works based on the application of these two proposals within the framework of general relativity. Whether all possible measures of GE available at this moment can be brought under a common roof in the context of general relativity, is yet to be ascertained. Even then several other studies conducted within the scope of general relativity could not be discussed here due to real life constraints. With the emergence of numerous models of alternative theories of gravity, the investigations of the measures of gravitational entropy in such models acquire even greater significance and require far careful examinations. A comprehensive account of such works is still in waiting.

\section*{Acknowledgments}
This work was initiated and completed during two successive visits in consecutive years to IUCAA, India, under the Associateship programme. SG gratefully acknowledges the warm hospitality and the facilities of work at IUCAA, and also thanks CSIR, Government of India for approving the major research project No. 03(1446)/18/EMR-II.


\begin{thebibliography}{}
\bibitem{1} We will use the hyphen in `space-time' as long as we discuss Tolman's work, but will do away with it afterwards.
\bibitem{2} The `hyphen' have been dropped for the remaining part of this article.
\end{thebibliography}
\end{document}